\documentclass[12pt]{iopart}
\usepackage{graphicx}
\usepackage{bm}
\usepackage{cite}
\usepackage{lscape}
\usepackage{hyperref}
\begin{document}

\title
    [Triple-$\alpha$ reaction rates below $T_9=3$ by a non-adiabatic three-body model]
    {Triple-$\alpha$ reaction rates below $T_9=3$ \\by a non-adiabatic three-body model}

\author{M. Katsuma$^{1,2}$}

\address{$^1$ Institut d'Astronomie et d'Astrophysique, Universit\'e Libre de Bruxelles, Brussels B-1050, Belgium}
\address{$^2$ Advanced Mathematical Institute, Osaka City University, Osaka 558-8585, Japan}
\ead{mkatsuma@gmail.com}
\vspace{10pt}
\begin{indented}
\item[]June 2025
\end{indented}

\begin{abstract}
The triple-$\alpha$ reaction from the ternary continuum states at off-resonant energies, $\alpha+\alpha+\alpha\rightarrow^{12}$C, remains an open question.
This direct process is scrutinized using a non-adiabatic Faddeev hyperspherical harmonics $R$-matrix expansion method, and the derived reaction rates are discussed.
After reviewing the model, the resultant photo-disintegration of $^{12}$C($2^+_1\rightarrow 0^+$) is shown to be much smaller than the values predicted by the adiabatic models for $0.15 \le E \le 0.35$ MeV.
Despite the large difference, the derived reaction rates are illustrated to be concordant with the current evaluated rates for $0.08 \le T_9 \le 3$.
The difference below $E=0.20$ MeV can be seen in the rates for $T_9 \le 0.07$.
In comparison with the calculations, the rates are found to be reduced by a factor of 10$^{-4}$ at $T_9=0.05$, because of an accurate description for $^8$Be break-up.
Uncertainties of the rates are also estimated by examining the sensitivity to the 3$\alpha$ potentials.
By introducing three-body $S$-factors and a resonant term, the present rates are expressed in an analytic form, and they are provided in a tabular form for astrophysical applications.
To update the evaluated rates, non-resonant sequential process between $\alpha+^8$Be could be removed.
The astrophysical impact is not expected to be large, although the rates are reduced around $T_9=0.05$.
\end{abstract}

\vspace{2pc}

\maketitle

\section{Introduction}\label{sec:1}
The triple-$\alpha$ reaction plays a crucial role in stellar evolution and concomitant nucleosynthesis \cite{Hoy54,Sal52}.
This reaction, followed by $^{12}$C($\alpha$,$\gamma$)$^{16}$O \cite{Kat08}, controls C/O ratio at the end of helium burning phase in stars, which determine the fate of massive stars up to their supernova explosion.
Whereas the $^{12}$C($\alpha$,$\gamma$)$^{16}$O reaction still has significant uncertainties of cross sections, the triple-$\alpha$ reaction is currently well understood through the experimental studies of the 0$^+_2$ state in $^{12}$C ($E=0.379$ MeV), e.g.~\cite{Nacre,Smi17,Del17}.
That is, the triple-$\alpha$ reaction rates at helium burning temperatures, $T_9 > 0.1$, have been determined relatively well with the sequential process via the narrow resonances: $\alpha+\alpha \rightarrow ^8$Be(0$^+_1$); $\alpha+^8$Be $\rightarrow ^{12}$C(0$^+_2$) \cite{Nacre,CF88}.
$T_9$ is temperature in units of 10$^9$ K; $E$ is the center-of-mass energy to the 3$\alpha$ threshold in $^{12}$C.

Apart from the sequential process, the triple-$\alpha$ reaction from the ternary continuum states is referred to as the direct triple-$\alpha$ process: $\alpha+\alpha+\alpha \rightarrow ^{12}$C.
This process is generally expected to be very slow because the three $\alpha$-particles must almost simultaneously collide and fuse into a $^{12}$C nucleus.
Thus, it is neglected or treated in certain approximations.
In NACRE \cite{Nacre}, $^8$Be is assumed to be bound as a particle, and the reaction rates have been estimated using an improved model based on the pioneering works of \cite{Nom85,Lan86}.
To ponder 3$\alpha$ continuum states more theoretically, formulae in hyperspherical coordinates have been adopted in e.g.~\cite{Fed96}, and recently the Coulomb modified Faddeev (CMF) method \cite{Ish13,Ish14,Ish16} and adiabatic channel function (ACF) expansion method \cite{Sun16} may have achieved successful progress quantitatively.
However, the direct triple-$\alpha$ process, which may dominate the rates at low temperatures ($T_9 < 0.1$), still seems to remain an open question.
Especially, the direct process corresponds to the non-resonant component in \cite{Nom85,Lan86}, and it is expected to play an important role in astrophysical sites with extreme conditions, such as novae, x-ray bursts and type-Ia supernovae.
The direct process, when enhanced by a high density environment, generates large amounts of heat, and the resulting high temperature states may influence the nucleosynthesis of the hot CNO cycle, rp-process and $\alpha$p-process \cite{Wal81,Sch06}.
Therefore, the theoretical studies are required to estimate the reaction rates at low temperatures more precisely.

In fact, CMF and ACF postulate adiabatic conditions using a cutoff procedure \cite{Ish13} and adiabatic channel functions \cite{Sun16}.
``Adiabatic'' means an approximation assuming a large difference of velocity in different coordinates describing the $\alpha$-particles.
Let us consider $a+A(b+c)$ scattering as an example.
$a$, $b$, $c$ are particles. $A$ consists of $b$ and $c$.
Here, we may perhaps assume that the relative motion between $b+c$ is very slow, compared with the velocity between $a+A$.
In this situation, incoming-$a$ can see a pair of $b+c$ keeping their distance, and the scattering amplitudes are described separately for a fixed distance of the pair.
The final cross sections are obtained by integrating the components over the distance.
The adiabatic feature is constructed under the constrained coordinates \cite{Sat,Tho09}.
One of the features of adiabatic approximation is that the amplitudes are given separately for each value of the constrained coordinates.

In CMF, the relative motion between $b+c$ is assumed to be slow, and $a$ is treated as a spectator.
$a+A$ is given adiabatically by the regular Coulomb wavefunctions ((26) of \cite{Ish13}) because spectators do not actively participate in reactions.
The relative motion between $b+c$ is described by solving an ordinary differential equation with coupling terms.
The break-up amplitudes are described separately for a conjugate momentum of the distance of $b+c$ ((35) of \cite{Ish13}).
The cross sections are obtained by integrating them in (13) of \cite{Ish13}.
CMF is executed including the channels arising from the cyclic permutation.
Accordingly, $b$ and $c$ are also treated as the spectator in $b+B(c+a)$ and $c+C(a+b)$.
Due to the symmetric formula, the hyper-radii defined by (10) of \cite{Ish13} are essentially the constrained coordinates.
CMF originally worked by describing three-body bound and continuum states below the three-body breakup threshold, such as p+d reactions.
The wavefunction of the pair, e.g. $b+c$ in $A$, is assumed to be localized within a certain distance.
Above the three-body breakup threshold, CMF collapsed in numerical calculations due to some difficulties in taking account of the long-range Coulomb force.
As an approximation, CMF runs stably in break-up channels by using the cutoff factor in integral kernels, as if it did below the three-body breakup threshold.

In ACF, the hyper-radial motion (the variation in the global size) is assumed to be slow.
ACF is based on the adiabatic potentials generated from the fixed-radius Schr\"odinger equation ((20) of \cite{Sun16}), and the transition amplitudes are described separately for a fixed hyper-radius through adiabatic channel functions.
ACF is governed by the adiabatic Hamiltonian, and it derives an adiabatic solution in convergence.
The derived adiabatic potentials show that the internal structure of $^{12}$C is described by $\alpha$+$^8$Be, and that the external region is expanded by the 3$\alpha$ configuration.
At low energies, the contribution from large hyper-radii becomes important in the strength function.
Thus, the photo-disintegration cross sections for $^{12}$C(2$^+_1\rightarrow$ 0$^+$) appear to be described by the non-resonant 3$\alpha$ component.
In \cite{Sun16}, the calculated cross sections have a kink around $E=0.13$ MeV, and this has been interpreted as the boundary between the non-resonant regime and the resonant regime.

To describe the ternary continuum more rigorously, a group of \cite{Ngu12,Ngu13} has adopted the Faddeev hyperspherical harmonics $R$-matrix (HHR) expansion method.
HHR is a non-adiabatic model, which is based on $R$-matrix theory \cite{Lan58} describing the 3$\alpha$ bound and continuum states quantum-mechanically.
However, it has been reported that the calculated photo-disintegration cross sections of HHR have a large discrepancy from those of CMF and ACF at off-resonant energies. (e.g. figure~4 of \cite{Sun16}).
Owing to the discrepancy, HHR makes the large difference in the reaction rates at low temperatures.
On the other hand, the derived rates are consistent with the NACRE rates above $T_9=0.07$.
In other words, HHR may give the reasonable values of cross sections at the Gamow energy-window energies \cite{Nacre}, $E > 0.20$ MeV.
What do you think is the difference between the calculated cross sections at off-resonant energies?
It should be noted that CMF and ACF are the adiabatic models, which have restricted conditions of internal motion.
In reality, the relative motion among $\alpha$-particles ought to vary more flexibly during the collision.

In the present article, I discuss the direct triple-$\alpha$ process and derived reaction rates by using a non-adiabatic Faddeev HHR expansion method \cite{Kat23,Ngu12,Ngu13,Tho09,Des10,Des}.
I illustrate the calculated photo-disintegration cross sections of $^{12}$C($2^+_1 \rightarrow 0^+$).
At the same time, I discuss the difference between adiabatic and non-adiabatic approaches.
I show that the current evaluated rates are reduced by a factor of 10$^{-4}$ at $T_9 = 0.05$, because of an accurate description for $^8$Be break-up.
In addition, uncertainties of the rates are estimated by examining the sensitivity to the 3$\alpha$ potentials.
The calculated results shown here are independent of the preceding study with HHR \cite{Ngu12,Ngu13}.
To avoid confusion, my calculation is referred to as HHR$^\ast$.

This article consists of six sections.
The following section reviews the Faddeev HHR$^\ast$ expansion method used in the present study.
\Sref{sec:3} describes a formula of reaction rates for the triple-$\alpha$ reaction by introducing a three-body type of $S$-factors.
The interaction potentials for $\alpha$+$\alpha$ and 3$\alpha$ are examined, and the density distribution functions of the states in $^{12}$C are illustrated in \sref{sec:4}.
The calculated triple-$\alpha$ reaction rates are discussed in \sref{sec:5}, with illustrating the photo-disintegration of $^{12}$C.
The derived reaction rates are expressed in an analytic form, and they are listed in a numerical table for astrophysical applications.
They are also converted into REACLIB format \cite{Reaclib}.
For a brief assessment of the rates, I examine the ignition density \cite{Nom85,Lan86} of accreting white dwarfs.
The summary is in \sref{sec:6}.

\section{Non-adiabatic Faddeev HHR$^\ast$ expansion method}\label{sec:2}

In this section, I first introduce the three-body Schr\"odinger equation and define basis functions in hyperspherical coordinates, and then expound $3\alpha$ symmetrization and describe coupling potentials in the derived coupled-channel (CC) equations.
After recalling eigenvalue equations, I review the $R$-matrix expansion method for 3$\alpha$ continuum states.
Finally, $R$-matrix propagation for numerical solutions in the external region is explained.

\subsection{Faddeev hyperspherical harmonics (HH) expansion}\label{sec:2.1}

The triple-$\alpha$ system satisfies the three-body Schr\"odinger equation,
\begin{eqnarray}
  (\hat{H}_{3\alpha} -E)\Psi &=& 0.
  \label{eq:3b}
\end{eqnarray}
$\hat{H}_{3\alpha}$ is three-body Hamiltonian, $\hat{H}_{3\alpha}=\hat{T}+\hat{V}_{12}+\hat{V}_{23}+\hat{V}_{31}+\hat{V}_{3\alpha}$.
$\hat{T}$ is the kinetic energy operator.
$\hat{V}_{ij}$ is interaction between two $\alpha$-particles, and $\hat{V}_{3\alpha}$ is 3$\alpha$ interaction.
This equation is expressed as the so-called Faddeev equations, consisting of three components:
\begin{eqnarray}
  & &\left\{\begin{array}{ccc}
  (\,\hat{T}_1 +\hat{V}_{23}+\hat{V}_{3\alpha}-E \,)\, \Psi_{(1)} &=& -\hat{V}_{23} \Psi_{(2)} -\hat{V}_{23} \Psi_{(3)} \\
  (\,\hat{T}_2 +\hat{V}_{31}+\hat{V}_{3\alpha}-E \,)\, \Psi_{(2)} &=& -\hat{V}_{31} \Psi_{(3)} -\hat{V}_{31} \Psi_{(1)} \\
  (\,\hat{T}_3 +\hat{V}_{12}+\hat{V}_{3\alpha}-E \,)\, \Psi_{(3)} &=& -\hat{V}_{12} \Psi_{(1)} -\hat{V}_{12} \Psi_{(2)} \\
  \end{array}\right.
  \label{eq:Faddeev}
\end{eqnarray}
\begin{figure}[t] 
  \begin{center}
    \includegraphics[width=0.75\linewidth]{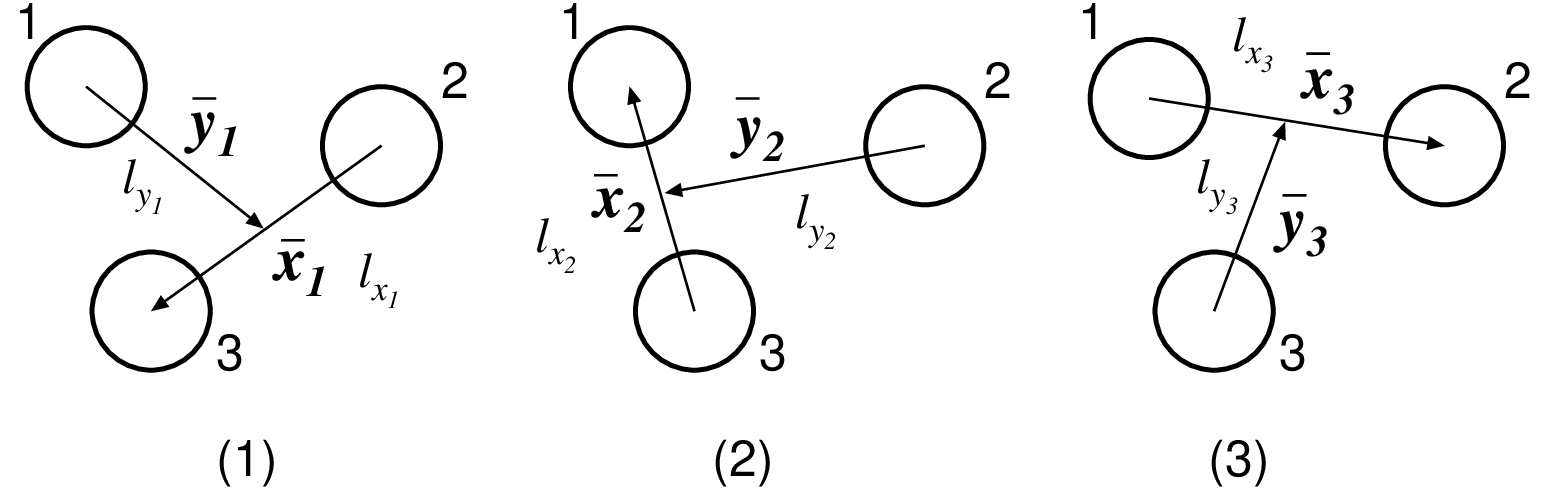}
  \end{center}
  \caption{\label{fig:jacobi}
Jacobi coordinates, used in the present article.
  }
\end{figure}
The total wavefunction is a sum of the components, $\Psi = \Psi_{(1)}+\Psi_{(2)}+\Psi_{(3)}$.
Jacobi coordinates ($\bar{\bm{x}}_i$, $\bar{\bm{y}}_i$) are defined as y-type ($i=1,2$) and T-type ($i=3$) in \fref{fig:jacobi}.
$\hat{T}_i$ is the kinetic energy operator for the Jacobi coordinate set $i$, $\hat{T}=\hat{T}_i\equiv\hat{T}_{\bar{x}_i}+\hat{T}_{\bar{y}_i}$.
Due to the symmetric 3$\alpha$ system, the coupled-equations of (\ref{eq:Faddeev}) are reduced to the single component with the sum of three pairwise interaction and 3$\alpha$ force, $(\,\hat{T}_3 +\hat{V}-E\,)\,\Psi_{(3)} = 0$, as described at the end of this subsection.
Let me rewrite the third component in
\begin{eqnarray}
  &&\hspace{-5mm}
  \left(
    -\frac{\hbar^2}{2\mu_{12}} \bm{\nabla}_{\bar{x}_3}^2
    -\frac{\hbar^2}{2\mu_{(12)3}} \bm{\nabla}_{\bar{y}_3}^2
    + \hat{V} \right) \Psi_{lm} = E \,\Psi_{lm},
  \label{eq:schrod}
  \\
  &&\hspace{-5mm}
  \hat{V} = \mathop{\sum}_{i<j}^3 \hat{V}_{ij} + \hat{V}_{3\alpha}.
\end{eqnarray}
$l$ is spin of the states in $^{12}$C.
$m$ is the projection of $l$.
$\mu_{12}$ and $\mu_{(12)3}$ are the reduced mass for $\alpha$+$\alpha$ and $\alpha$+$^8$Be.
If Jacobi coordinates are scaled as $\bm{x}_3\equiv\sqrt{\mu_{12}/m_{\rm N}}\, \bm{\bar{x}}_3$ and $\bm{y}_3\equiv\sqrt{\mu_{(12)3}/m_{\rm N} }\,\bm{\bar{y}}_3$, (\ref{eq:schrod}) becomes
\begin{eqnarray}
  \left[ -\frac{\hbar^2}{2m_{\rm N}} (\bm{\nabla}_{x_i}^2+\bm{\nabla}_{y_i}^2) + \hat{V} \right] \Psi_{lm} &=& E \,\Psi_{lm},
  \label{eq:scaled-eqs}
\end{eqnarray}
where $m_{\rm N}$ is nucleon mass.
From the scaled Jacobi coordinates, the hyperspherical coordinates $(\rho,\theta_i,\hat{\bm{x}}_i,\hat{\bm{y}}_i)$ are defined by
\numparts
\begin{eqnarray}
  \rho^2 &=& x^2_i+y^2_i,
  \label{eq:hyperradius} \\
  \theta_i &=& \arctan\left(\frac{x_i}{y_i}\right).
  \label{eq:hyperangle}
\end{eqnarray}
\endnumparts
$\rho$ and $\theta_i$ are the hyper-radius and hyperangle, respectively. (e.g.~\cite{Tho09})
When $\Psi_{lm}$ is separated into the hyper-radial part and hyperangle part (as (\ref{eq:basis})), equation~(\ref{eq:scaled-eqs}) is transformed into a similar form of ordinary CC equations with $L=K+3/2$ for inelastic scattering (e.g.~\cite{Kat02,Sat}),
\begin{eqnarray}
  \left[\, T_\gamma +U^l_{\gamma \gamma}(\rho)-\epsilon \,\right] \chi^l_{\gamma}(\rho)
  &=&
  -\mathop{\sum}_{\gamma^\prime \ne\gamma} U^l_{\gamma^\prime \gamma}(\rho) \chi^l_{\gamma^\prime}(\rho),
  \label{eq:hscc}
\end{eqnarray}
where
\begin{eqnarray}
  T_\gamma   &=& \frac{d^2}{d\rho^2} -\frac{(K+3/2)(K+5/2)}{\rho^2},
  \label{eq:Tg}
  \\
  U^l_{\gamma^\prime \gamma}(\rho) &=& -\frac{2m_{\rm N}}{\hbar^2} V^l_{\gamma^\prime\gamma}(\rho),
  \label{eq:ccp}
  \\
  \epsilon &=& -\frac{2m_{\rm N}}{\hbar^2}E.
\end{eqnarray}
$\chi^l_{\gamma}(\rho)$ are hyper-radial wavefunctions.
The hyperangle part is solved with Jacobi polynomials.
$V^l_{\gamma^\prime \gamma}$ are coupling potentials.
$K$ are hyperangular momenta.
$\gamma$ is a label of channels, $\gamma=(K,l_{x_3},l_{y_3})$.
$l_{x_3}$ and $l_{y_3}$ are internal angular momenta in Jacobi coordinates.
$l_{x_3}$ is only even.
This condition of the partial wave ensures the symmetrization of the system.
$l_{y_3}$ is also even for 0$^+$ and 2$^+$ because the parity of the 3$\alpha$ system is given by $\pi=(-)^{l}=(-)^{l_{x_3}+l_{y_3}}$.
The $K=2$ channel for 0$^+$ is vanished due to the symmetric 3$\alpha$ system.

The basis functions are defined by
\begin{eqnarray}
  \Psi_{lm} &=&
  \rho^{-5/2} \mathop{\sum}_{\gamma}\chi^l_{\gamma}(\rho)\,\Phi^\gamma_{lm}(\Omega_5^{(3)}),
  \label{eq:basis}
  \\
  \chi^l_{\gamma}(\rho) &=& \mathop{\sum}_n \bar{c}^n_\gamma w_n^\gamma(\rho),
  \label{eq:chi}
\end{eqnarray}
where $\Omega_5^{(3)} \equiv (\theta_3, \hat{\bm{x}}_3, \hat{\bm{y}}_3)$.
$n$ is the number of radial nodes ($n=0,1,\cdots$).
$K$ in $\gamma$ runs up to $K=26$ for 0$^+$ \cite{Ngu13} and $K=12$ for 2$^+$ \cite{Ngu13}.
The internal angular momenta up to $l_{x_3}=12$ and $l_{y_3}=12$ are taken into account.
$\bar{c}^n_\gamma$ are the coefficients of expansion.
$w_n^\gamma(\rho)$ are eigenfunctions, satisfying the following equation with eigen-energies $\epsilon_{n\gamma}$:
\begin{eqnarray}
  \left[\, T_\gamma +U^l_{\gamma \gamma}(\rho)-\epsilon_{n\gamma} \,\right] w_n^\gamma(\rho) &=& 0.
\end{eqnarray}
$w_n^\gamma(\rho)$ are assumed to have a fixed logarithmic derivative $b_c \equiv d \ln w_n^\gamma(\rho)/d\rho = 0$ at $\rho=a_c$, i.e., they ensure that the $b_c=0$ formula of $R$-matrix theory \cite{Tho09,Des,Lan58} can be used.
The hyperharmonic functions are defined by
\begin{eqnarray}
  \Phi^\gamma_{lm}(\Omega_5^{(3)}) &=&
  \phi_K^{l_{x_3}l_{y_3}}(\theta_3)\left[{\cal Y}_{l_{x_3}}(\hat{\bm{x}}_3)\otimes {\cal Y}_{l_{y_3}}(\hat{\bm{y}}_3) \right]_{lm},
  \label{eq:hsang}
\end{eqnarray}
where 
${\cal Y}_{lm}\equiv i^l Y_{lm}$.
$Y_{lm}$ are spherical harmonics.
$\phi_K^{l_x l_y}$ are defined by
\begin{eqnarray}
  \phi_K^{l_{x_3}l_{y_3}}(\theta_3)
  &=&
  N_{n_\theta}^{l_{x_3}l_{y_3}}(\sin\theta_3)^{l_{x_3}} (\cos\theta_3)^{l_{y_3}} {\cal G}_{n_\theta}(n_a,n_b;\sin^2\theta_3),
  \label{eq:phi}
  \\
  N_{n_\theta}^{l_{x_3}l_{y_3}}
  &=&
  \left[\frac{2(n_a+2n_\theta)\Gamma(n_a+n_\theta)\Gamma(n_b+n_\theta)}{n_{\theta}!\Gamma(n_\theta+l_{y_3}+3/2)}\right]^\frac{1}{2} 
  \frac{1}{\Gamma(n_b)}.
\end{eqnarray}
${\cal G}_{n_\theta}$ are Jacobi polynomials, $n_\theta=\frac{1}{2}(K-l_{x_3}-l_{y_3 })$, $n_a=l_{x_3}+l_{y_3}+2$, $n_b=l_{x_3}+3/2$.
$\Gamma$ are gamma functions.
The values of $K$ are only even because $(l_{x_3}+l_{y_3})$ are even.
$\Phi^\gamma_{lm}(\Omega_5^{(3)})$ are independent of $\rho$, and they are different from the adiabatic channel functions in \cite{Sun16}.
The basis functions satisfy the following orthogonal conditions in a box with the finite interval:
\begin{eqnarray}
  \int_0^{a_c} w^\gamma_{n^\prime}(\rho) w^\gamma_n(\rho) d\rho &=& \delta_{n^\prime n},
  \\
  \int \Phi^{\gamma^\prime \,\ast}_{l^\prime m^\prime }(\Omega_5^{(3)})\,\Phi^\gamma_{lm}(\Omega_5^{(3)}) \,d\Omega_5^{(3)}
  &=&
  \delta_{\gamma^\prime \gamma} \delta_{l^\prime l} \delta_{m^\prime m}.
  \label{eq:oth5}
\end{eqnarray}

$\Phi^\gamma_{lm}(\Omega_5^{(3)})$ are expanded by the hyperharmonic functions in the y-type coordinates of \fref{fig:jacobi}, as follows:
\begin{eqnarray}
  \Phi^{K l_{x_3} l_{y_3}}_{lm}(\Omega_5^{(3)})
  &=& \mathop{\sum}_{l_{x_i} l_{y_i}} \langle l_{x_i}l_{y_i}| l_{x_3}l_{y_3} \rangle_{Kl} \Phi^{K l_{x_i} l_{y_i}}_{lm}(\Omega_5^{(i)}),
  \label{eq:uphi}
\end{eqnarray}
where $\langle l_{x_i} l_{y_i}| l_{x_3} l_{y_3} \rangle_{Kl}$ are the Raynal-Revai (RR) coefficients \cite{Ray70}.
$K$, $l$, $m$ are conserved in this unitary transformation.
The RR coefficients are used for convenience when nuclear potentials depend on $l_{x_i}$ (see (\ref{eq:ccpot})).

$w_n^\gamma(\rho)$ are expanded by a set of arbitrary orthonormal functions $\varphi_{n^\prime}^K(\rho)$, as follows:
\begin{eqnarray}
  w_n^\gamma(\rho) &=& \mathop{\sum}_{n^\prime=0}^{n_\rho-1} d_{\gamma n}^{n^\prime} \varphi^K_{n^\prime}(\rho),
  \label{eq:varphi}
\end{eqnarray}
where $d_{\gamma n}^{n^\prime}$ are the coefficients of expansion.
($n_\rho-1$) is the maximum number of radial nodes.
If the coefficients are rewritten as $c^{n^\prime}_\gamma=\mathop{\sum}_n \bar{c}^n_\gamma d_{\gamma n}^{n^\prime}$, $w_n^\gamma(\rho)$ in the formula can be replaced with $\varphi_n^K(\rho)$.

The wavefunctions in (\ref{eq:schrod}) and (\ref{eq:hscc}) are therefore expanded as
\begin{eqnarray}
  \Psi_{lm} &=& \rho^{-5/2}\mathop{\sum}_{\gamma n} c^n_\gamma 
  \varphi^K_n(\rho) \Phi^\gamma_{lm}(\Omega_5^{(3)}),
  \label{eq:hpex}
  \\
  \chi^l_{\gamma}(\rho) &=& \mathop{\sum}_{n=0}^{n_\rho-1} c^n_\gamma \varphi^K_n(\rho).
  \label{eq:hpw}
\end{eqnarray}
$\Psi_{lm}$ is normalized as
\begin{equation}
  \int \big|\,\Psi_{lm}\,\big|^2 \rho^5 d\rho d\Omega_5^{(3)}
  =
  \int \big|\,\Psi_{lm}\,\big|^2 d\bm{x}_3 d\bm{y}_3
  =
  {\cal J}^3 \int \big|\,\Psi_{lm}\,\big|^2 d\bar{\bm{x}}_3 d\bar{\bm{y}}_3  
  = 1,
\end{equation}
where ${\cal J}=\sqrt{\mu_{12}\mu_{(12)3}/m_{\rm N}^2}=4/\sqrt{3}$.
If a large number of basis functions are used in calculations, the final results are independent of a choice of $\varphi^K_n(\rho)$.

In HHR$^\ast$, the harmonic oscillator wavefunctions in hyperspherical coordinates are adopted as the orthonormal functions,
\begin{eqnarray}
  \varphi_n^K(\rho)\! &=& \!\left[\frac{2\nu n!}{(n+K+2)!}\right]^{\frac{1}{2}}\!\!
  (\nu\rho)^{K+\frac{5}{2}} \exp\Big( -\frac{1}{2} (\nu\rho)^2\Big) L^{(K+2)}_n((\nu\rho)^2),
  \label{eq:gwf0}
\end{eqnarray}
where $L_n^{(K+2)}$ are associated Laguerre polynomials.
$\nu$ is a harmonic oscillator range.
$\nu=0.23$ fm$^{-1}$ and $n_\rho=160$ are used to make expansion efficiently.
The derivatives of (\ref{eq:gwf0}) are
\begin{eqnarray}
  \frac{d\varphi_n^K(\rho)}{d\rho} &=&
  \frac{1}{\rho}\,\Big[ \big(K+\frac{5}{2}+2n-(\nu\rho)^2\big) \,\varphi_n^K(\rho)
        -2\sqrt{n(n+K+2)}\,\varphi_{n-1}^K(\rho) \Big].
  \label{eq:dgwf0}
  \nonumber \\
  && 
\end{eqnarray}

In the last part of this subsection, I describe the symmetric properties of the 3$\alpha$ wavefunction \cite{Tho09,Ngu12,Ngu13}.
With respect to the exchange of $\alpha$-particles $1\leftrightarrow 2$, $\Psi_{(1)}$ has a relation as
\numparts
\begin{eqnarray}
  \begin{array}{ccccc}
    \hat{\cal P}_{12}\Psi_{(1)}&=&    \Psi_{(2)}    &=& (-)^{l_{x_1}}\Psi_{(1)},
  \end{array}
\end{eqnarray}
where $\hat{\cal P}_{ij}$ is the exchange operator between two $\alpha$-particles.
The phase comes out from ${\cal Y}_{l_{x_i}}(-\hat{\bm{x}}_i)=(-)^{l_{x_i}} {\cal Y}_{l_{x_i}}(\hat{\bm{x}}_i)$ in (\ref{eq:hsang}).
$\Psi_{(2)}$ and $\Psi_{(3)}$ have
\begin{eqnarray}
  \begin{array}{ccccc}
    \hat{\cal P}_{12}\Psi_{(2)}&=&    \Psi_{(1)}    &=& (-)^{l_{x_2}}\Psi_{(2)},\\
    \hat{\cal P}_{12}\Psi_{(3)}&=& \Psi^\prime_{(3)} &=& (-)^{l_{x_3}}\Psi_{(3)}.\\
  \end{array}
\end{eqnarray}
Likewise, $2\leftrightarrow 3$:
\begin{eqnarray}
  \begin{array}{ccccc}
    \hat{\cal P}_{23}\Psi_{(1)}&=& \Psi^\prime_{(1)} &=& (-)^{l_{x_1}}\Psi_{(1)},\\
    \hat{\cal P}_{23}\Psi_{(2)}&=&    \Psi_{(3)}    &=& (-)^{l_{x_2}}\Psi_{(2)},\\
    \hat{\cal P}_{23}\Psi_{(3)}&=&    \Psi_{(2)}    &=& (-)^{l_{x_3}}\Psi_{(3)}.\\
  \end{array}
\end{eqnarray}
$3\leftrightarrow 1$:
\begin{eqnarray}
  \begin{array}{ccccc}
    \hat{\cal P}_{31}\Psi_{(1)}&=&    \Psi_{(3)}    &=& (-)^{l_{x_1}}\Psi_{(1)},\\
    \hat{\cal P}_{31}\Psi_{(2)}&=& \Psi^\prime_{(2)} &=& (-)^{l_{x_2}}\Psi_{(2)},\\
    \hat{\cal P}_{31}\Psi_{(3)}&=&    \Psi_{(1)}    &=& (-)^{l_{x_3}}\Psi_{(3)}.\\
  \end{array}
\end{eqnarray}
\endnumparts
Therefore, the exchange between any pair of $\alpha$-particles to the total wavefunctions is obtained as
\begin{eqnarray}
  \hat{\cal P} \Psi &=& (-)^{l_{x_1}}\Psi_{(1)}+(-)^{l_{x_2}}\Psi_{(2)}+(-)^{l_{x_3}}\Psi_{(3)},
\end{eqnarray}
where $\hat{\cal P}=(\hat{\cal P}_{12}+\hat{\cal P}_{23}+\hat{\cal P}_{31})/3$. 
If $l_{x_1}$, $l_{x_2}$, and $l_{x_3}$ are even, I find
\begin{eqnarray}
  \hat{\cal P} \Psi &=& \Psi.
\end{eqnarray}
Under this constraint on the partial waves, the right hand of the first component of (\ref{eq:Faddeev}) becomes
\numparts
\begin{eqnarray}
  \hat{V}_{23} \Psi_{(2)} &=& \hat{V}_{31} \Psi_{(1)},
  \label{eq:fdv1}
  \\
  \hat{V}_{23} \Psi_{(3)} &=& \hat{V}_{12} \Psi_{(1)}.
  \label{eq:fdv2}
\end{eqnarray}
\endnumparts
The exchange of 1 $\leftrightarrow$ 2 (3 $\leftrightarrow$ 1) is used in (\ref{eq:fdv1}) ((\ref{eq:fdv2})).
Thus, the first component of (\ref{eq:Faddeev}) is rewritten as
\begin{eqnarray}
  [\,\hat{T}_1 +(\hat{V}_{12}+\hat{V}_{23}+\hat{V}_{31}) +\hat{V}_{3\alpha}-E \,]\, \Psi_{(1)} &=& 0.
  \label{eq:faddeev-1}
\end{eqnarray}
The same transformation can be performed on the second and third Faddeev components.
Consequently, three sets of the independent CC equations are obtained in the same expression.
In other words, this 3$\alpha$ symmetrization efficiently reduces the number of channels included in (\ref{eq:Faddeev}).
Note that $\hat{\cal P}$ and $\hat{\cal P}_{ij}$ are the boson exchange operators when $l_{x_i}$ are even.

\subsection{Coupling potentials}\label{sec:2.2}

The coupling potentials in (\ref{eq:ccp}) are given in
\begin{eqnarray}
  V^l_{\gamma^\prime\gamma}(\rho)
  &=&
  \int \Phi^{\gamma^\prime \,\ast}_{lm}(\Omega_5^{(3)}) \hat{V} \,\Phi^\gamma_{lm}(\Omega_5^{(3)}) \,d\Omega_5^{(3)}
  \nonumber \\
  &=&
  V^l_{3\alpha}(\rho) \delta_{\gamma^\prime\gamma} +
  \delta_{l_{x_3}^\prime l_{x_3}}\delta_{l_{y_3}^\prime l_{y_3}} F^{l_{x_3} l_{y_3}}_{K^\prime K}(\rho)
  \nonumber \\
  &+&
  \mathop{\sum}_{l_{x_1} l_{y_1}} \langle l_{x_1} l_{y_1} | l^\prime_{x_3} l^\prime_{y_3} \rangle_{K^\prime l}
  \langle l_{x_1} l_{y_1} | l_{x_3} l_{y_3} \rangle_{Kl} F^{l_{x_1} l_{y_1}}_{K^\prime K}(\rho)
  \nonumber \\                             
  &+&
  \mathop{\sum}_{l_{x_2} l_{y_2}} \langle l_{x_2} l_{y_2} | l^\prime_{x_3} l^\prime_{y_3} \rangle_{K^\prime l}
  \langle l_{x_2} l_{y_2} | l_{x_3} l_{y_3} \rangle_{Kl} F^{l_{x_2} l_{y_2}}_{K^\prime K}(\rho),
  \label{eq:ccpot}
\end{eqnarray}
where
\begin{eqnarray}
 F^{l_{x_i} l_{y_i}}_{K^\prime K}(\rho) &=&
  \int \phi^{l_{x_i}l_{y_i}}_{K^\prime}(\theta_i)\,V^{l_{x_i}}_{2\alpha}(\bar{x}_i)\,\phi^{l_{x_i}l_{y_i}}_K (\theta_i)
  \cos^2\theta_i \sin^2\theta_i d\theta_i.
  \label{eq:fkk}
\end{eqnarray}
$V^{l_{x_i}}_{2\alpha}$ and $V^l_{3\alpha}$ are $\alpha$+$\alpha$ and 3$\alpha$ interaction potentials, respectively.
$V^{l_{x_i}}_{2\alpha}$ consists of two parts, nuclear and Coulomb potentials, $V^{l_{x_i}}_{2\alpha}=V^{l_{x_i}}_{\rm N}+V_{\rm C}$.

Both diagonal and non-diagonal components of (\ref{eq:ccpot}) vary as $Z^l_{\gamma^\prime \gamma}e^2/\rho$ in the asymptotic region \cite{Ngu13,Des10,Vas12}.
In HHR$^\ast$, $Z^l_{\gamma^\prime \gamma}$ are given as
\begin{eqnarray}
  Z^l_{\gamma^\prime \gamma}
  &=&
  4\sqrt{2} \Big( \delta_{l_{x_3}^\prime l_{x_3}}\delta_{l_{y_3}^\prime l_{y_3}} F^{l_{x_3} l_{y_3}}_{{\rm (C)} K^\prime K}
  \nonumber \\
  &+&
  \mathop{\sum}_{l_{x_1} l_{y_1}} \langle l_{x_1} l_{y_1} | l^\prime_{x_3} l^\prime_{y_3} \rangle_{K^\prime l}
  \langle l_{x_1} l_{y_1} | l_{x_3} l_{y_3} \rangle_{Kl}
  F^{l_{x_1} l_{y_1}}_{{\rm (C)} K^\prime K}
  \nonumber \\
  &+&
  \mathop{\sum}_{l_{x_2} l_{y_2}} \langle l_{x_2} l_{y_2} | l^\prime_{x_3} l^\prime_{y_3} \rangle_{K^\prime l}
  \langle l_{x_2} l_{y_2} | l_{x_3} l_{y_3} \rangle_{Kl}
  F^{l_{x_2} l_{y_2}}_{{\rm (C)} K^\prime K} \Big),
  \nonumber
\\
  &=&
  4\sqrt{2}\Big(
  \delta_{l^\prime_{x_3} l_{x_3}} \delta_{l^\prime_{y_3} l_{y_3}} F^{l_{x_3}l_{y_3}}_{{\rm (C)}K^\prime K}
  + \mathop{\sum}_t {\cal I}^{l\,l_{y_3}^\prime l_{x_3}^\prime}_{t l_{x_3} l_{y_3}}
  \nonumber\\
  &\cdot&
  \int
  \phi^{l^\prime_{x_3} l^\prime_{y_3} }_{K^\prime}(\theta_3)
  \bar{f}_t(\theta_3) 
  \phi^{l_{x_3}l_{y_3}}_K(\theta_3)
  \cos^2\theta_3 \sin^2\theta_3 d\theta_3 \Big),
  \label{eq:zeff}
\end{eqnarray}
where
\begin{eqnarray}
   F^{l_{x_i} l_{y_i}}_{{\rm (C)} K^\prime K} &=&
  \int \phi^{l_{x_i}l_{y_i}}_{K^\prime}(\theta_i)\,\frac{1}{\sin\theta_i}\,\phi^{l_{x_i}l_{y_i}}_K (\theta_i)
  \cos^2\theta_i \sin^2\theta_i d\theta_i,
  \label{eq:ffc}
  \\
  {\cal I}^{l\,l_{y_3}^\prime l_{x_3}^\prime}_{t l_{x_3} l_{y_3}}
  &=&
  i^{(l_{x_3}-l_{x_3}^\prime +t) + (l_{y_3}-l_{y_3}^\prime +t)} \,(-)^l\,
  \hat{l}_{x_3}^\prime \hat{l}_{y_3}^\prime \hat{l}_{x_3} \hat{l}_{y_3} (2t+1)
  \nonumber\\
  &\cdot&
  \left(\begin{array}{ccc}
    l_{x_3}^\prime & t & l_{x_3} \\
    0   &   0     &   0   \\
  \end{array}\right)
  \left(\begin{array}{ccc}
    l_{y_3}^\prime & t & l_{y_3} \\
    0   &   0     &   0   \\
  \end{array}\right)
  \left\{\begin{array}{ccc}
       l    & l_{y_3}^\prime & l_{x_3}^\prime \\
    t & l_{x_3}       & l_{y_3} \\
  \end{array}\right\},
  \\
  \bar{f}_t(\theta_3)
  &=&
  \int^1_{-1} \frac{P_{t}(z) dz}{\sqrt{\xi_x^2+\xi_y^2+2\xi_x\xi_y z}},
  \label{eq:ffc2}
\end{eqnarray}
$\xi_x=(1/2)\sin\theta_3$, $\xi_y=(\sqrt{3}/2)\cos\theta_3$, $\hat{l}=\sqrt{2l+1}$.
$P_t$ are Legendre polynomials.
$t$ is only even.
$(:\,:\,:)$ and $\{:\,:\,:\}$ are 3-J and 6-J symbols, respectively.
The Sommerfeld parameter of the $\gamma$ channel is defined as $\eta_{\gamma}\equiv Z^l_{\gamma\gamma}e^2/(\hbar v)$, $v=\sqrt{2E/m_{\rm N}}$.

\subsection{Solving hyper-radial equations in the interior region}\label{sec:2.3}

The CC equations of (\ref{eq:hscc}) are expressed by eigenvalue equations as
\begin{eqnarray}
  (\bm{T}+\bm{U})\bm{X}&=&\epsilon\bm{X}.
  \label{eq:cceqmat}
\end{eqnarray}
$\bm{T}$ and $\bm{U}$ are $(n_\gamma n_\rho)\times(n_\gamma n_\rho)$ matrices of the kinetic energy operator and interaction.
$n_\gamma$ is the number of channels.
$\bm{X}$ is a coefficient vector with ($n_\gamma n_\rho$) components.
The matrix elements of $\bm{T}$ and $\bm{U}$ are given in
\begin{eqnarray}  
  \langle  \varphi_{n^\prime}^K | T_\gamma | \varphi_n^K\rangle
  &=&
  \left\{\begin{array}{cc}
    -\sqrt{n(n+K+2)}\, \nu^2 & n^\prime=n-1 \\ \\
    -(2n+K+3)\, \nu^2 & n^\prime=n \\ \\
    -\sqrt{(n+1)(n+K+3)}\, \nu^2 & n^\prime=n+1 \\ \\
    0 & {\rm others}\\
    \end{array}\right.
  \\
  && \nonumber \\
  \langle  \varphi_{n^\prime}^{K^\prime} | U^l_{\gamma^\prime\gamma} | \varphi_n^K\rangle
  &=&
  \int_0^{a_c} \varphi_{n^\prime}^{K^\prime}(\rho) U^l_{\gamma^\prime\gamma}(\rho) \varphi_n^K(\rho) d\rho.
  \label{eq:int_u}
\end{eqnarray}

\Eref{eq:cceqmat} is solved using matrix diagonalization for $\bm{H}=\bm{T}+\bm{U}$:
\begin{eqnarray}
  {}^t\bm{W}\bm{H}\bm{W}
  &=& \left(\begin{array}{cccc}
    \epsilon_1 &     0      & \ldots  & 0 \\
        0      & \epsilon_2 & \ldots  & 0  \\
      \vdots   &  \vdots    & \ddots & \vdots \\
        0      &     0      & \ldots  & \epsilon_{(n_\gamma n_\rho)} \\
            \end{array}\right),
  \\
  \bm{W} &=& \left(\bm{X}_1,\bm{X}_2, \cdots, \bm{X}_{(n_\gamma n_\rho)}\right).
\end{eqnarray}
$\bm{W}$ is a unitary matrix.
The linearly-independent solutions are expressed with $\epsilon_i$ and $\bm{X}_i$.
The eigen-energies are given by
\begin{eqnarray}
  \tilde{E}(l^\pi_i) &=& -\frac{\hbar^2}{2m_{\rm N}}\epsilon_i.
  \label{eq:eigen-en}
\end{eqnarray}
$i$ is a label of the states, given in order of the excitation energy.
$i=1$ means the lowest energy of states with $l^\pi$.
$\tilde{E}(0^+_2)$ is equivalent to the formal energy of 0$^+_2$ in $R$-matrix theory \cite{Lan58}.
It is however considered as the observed energy $E(0^+_2)$ because width is very narrow.
$\tilde{E}(l^+_1)$ for 0$^+_1$ and 2$^+_1$ are also observables, $\tilde{E}(l^+_1)=E(l^+_1)$.
The corresponding eigenfunctions are
\begin{eqnarray}
  \Psi_{i, l m} &=&
  \rho^{-5/2} \mathop{\sum}_{\gamma n} c^n_{\gamma i} \varphi_n^K(\rho) \Phi^{\gamma}_{lm}(\Omega_5^{(3)}),
  \label{eq:eigen-fun-tot}
  \\
  \chi^{l}_{\gamma i}(\rho) &=& \mathop{\sum}_{n=0}^{n_\rho-1} c^n_{\gamma i} \varphi_n^K(\rho).
  \label{eq:eigen-fun}
\end{eqnarray}
For $E(l^\pi_i) < 0$, $\chi^{l}_{\gamma i}(\rho)$ are bound state wavefunctions, extrapolated as
\begin{eqnarray}
  \chi^{l}_{\gamma i}(\rho) \rightarrow \tilde{\cal N}^{l}_{\gamma i} W_{-\eta_\gamma, K+2}(2\kappa\rho),
  \label{eq:tail}
\end{eqnarray}
where $\kappa=\sqrt{\epsilon_i}$; $W_{-\eta_\gamma, K+2}$ are Whittaker functions; $\tilde{\cal N}^{l}_{\gamma i}$ are normalization constants.
The tail of 2$^+_1$ is extrapolated using (\ref{eq:tail}).
For $E(l^\pi_i) > 0$, $\chi^{l}_{\gamma i}(\rho)$ are resonant state wavefunctions.
If resonant states exist experimentally in low-lying levels, the calculated eigenstates may correspond to the experimental levels.
However, most of the eigenstates are mathematically orthogonal states that do not have any specific features in physics.
The continuum states with the scattering boundary condition are expressed by a linear-combination of the eigenfunctions.
I describe it in \sref{sec:2.4}.

To understand nuclear structure of the generated eigenstates, density distribution functions are defined by
\begin{eqnarray}
  {\cal P}_{l^\pi_i}(\bar{x}_3,\bar{y}_3) &\equiv& {\cal J}^3 \bar{x}^2_3\,\bar{y}^2_3 \int \big| \Psi_{i, lm} \big|^2 \,d\hat{\bm{x}}_3 d\hat{\bm{y}}_3.
  \label{eq:dens}
\end{eqnarray}
The density is normalized as $\int {\cal P}_{l^\pi_i}(\bar{x}_3,\bar{y}_3) d\bar{x}_3 d\bar{y}_3 = 1$.
The reduced width amplitude for $\alpha$+$^8$Be in 0$^+_2$ is defined as
\begin{eqnarray}
  \tilde{\gamma}_{\alpha+^8{\rm Be}}(\bar{y}_3)&\equiv& \sqrt{\frac{{\cal J}^3\hbar^2}{2\mu_{(12)3}\bar{y}_3}}
  \bar{y}_3 \int \Psi_{\rm ^8Be}(\bar{\bm{x}}_3) \Psi_{2,00}(\bar{\bm{x}}_3,\bar{\bm{y}}_3) d\bar{\bm{x}}_3 d\hat{\bm{y}}_3.
  \label{eq:gamma_alpha}
\end{eqnarray}
$\Psi_{\rm ^8Be}$ is the wavefunction of $^8$Be(0$^+_1$).
The $\alpha$+$^8$Be width is given as
\begin{eqnarray}
  \Gamma_{\alpha+^8{\rm Be}}(\bar{y}_3) &=& 2 \tilde{\gamma}^2_{\alpha+^8{\rm Be}}(\bar{y}_3) P_{\alpha, 0}(\bar{y}_3),
  \label{eq:wa8Be}
\end{eqnarray}
where $P_{\alpha, 0}$ is the penetration factor, calculated from Coulomb wavefunctions (e.g.~\cite{Lan58,Tho09,Des}).
The $\alpha$-decay width of $^{12}$C(0$^+_2$) is calculated from $\Gamma(0^+_2) = 3\times\Gamma_{\alpha+^8{\rm Be}}$ in the present article.
The dimensionless reduced width is 
\begin{eqnarray}
  \theta^2_{\alpha+^8{\rm Be}}(\bar{y}_3) &=& \tilde{\gamma}^2_{\alpha+^8{\rm Be}}(\bar{y}_3)/\gamma^2_{\rm W},
  \label{eq:arw}
\end{eqnarray}
$\gamma^2_{\rm W}=3\hbar^2/(2\mu_{(12)3}\bar{y}_3^2)$.
The root-mean-square (rms) radius is given by
\begin{eqnarray}
  R_{\rm rms}(l^\pi_i) &=& \sqrt{ \,\langle r_\alpha^2 \rangle +
    \frac{1}{12}\mathop{\sum}_\gamma \int \big|\,\chi^l_{\gamma i}(\rho)\,\big|^2 \rho^2 d\rho},
  \label{eq:rms}
\end{eqnarray}
where $\langle r^2_\alpha \rangle^{1/2} = 1.4735$ fm (\sref{sec:4.1}).
The monopole matrix element \cite{Fed96,Sun16} between 0$^+_2$ and 0$^+_1$ is defined as
\begin{eqnarray}
  M(E0;0^+_2\rightarrow 0^+_1) &=&
  \frac{e}{2}\mathop{\sum}_\gamma \int \chi^{0^+}_{\gamma 2}(\rho)\chi^{0^+}_{\gamma 1}(\rho)\rho^2 d\rho.
  \label{eq:ME0}
\end{eqnarray}

The internal structure of the bound 2$^+_1$ state may not be described by a pure 3$\alpha$ component \cite{Sun16}, and the actual wavefunction is assumed to be replaced with
\begin{eqnarray}
  \Psi_{2^+_1} &=& {\cal N}\Psi_{1,2m}.
  \label{eq:2+}
\end{eqnarray}
As assumed in HHR, CMF and ACF, the resultant transition strength is given by multiplying the calculated one by a factor of ${\cal N}^2$.
In HHR$^\ast$, ${\cal N}^2$ is obtained by normalizing the calculated $\gamma$ width of 0$^+_2$ to $3.9\pm0.39$ meV \cite{Kel17}.

\subsection{$R$-matrix expansion}\label{sec:2.4}

$R$-matrix theory was developed by \cite{Lan58}, and it has been widely used to evaluate the experimental data phenomenologically.
In a basic idea of theory, low-energy nuclear reactions are assumed to be describable by the wavefunction values at a surface of nuclei after penetrating the effective barrier made by nuclear, Coulomb and centrifugal potentials.
The barrier is approximately located at the reach of nuclear force between colliding nuclei, and its position is defined as a channel radius $\rho=a_c$.
A compound nucleus is assumed to be formed in a sphere with the radius $a_c$, and the wavefunction connects to all possible reaction channels at $\rho=a_c$.
In the phenomenological evaluation of nuclear data, probabilities to the respective reaction channels are parameterized as the resonance energies and widths.
As a consequence, the resonant states in the interior region do not need to be solved microscopically.
In the external region, the wavefunctions are given by Coulomb wavefunctions because nuclear interaction is negligible.
This method is the {\it phenomenological} $R$-matrix method \cite{Des10b}.

In another $R$-matrix method, internal nuclear structure is assumed to be expanded by a set of arbitrary orthonormal basis, where internal wavefunctions are calculated microscopically with a model Hamiltonian.
Resonances, transition probabilities and reaction rates are derived from them on the theoretical footing.
This microscopic approach is referred to as the {\it calculable} $R$-matrix method \cite{Des10b}.
HHR$^\ast$ is classified as the latter.
The final result is independent of $a_c$ if many basis functions are used.
In this method, $a_c$ is set for convenience, and it appears to be different from the above-mentioned phenomenological definition, so that $a_c$ is referred to as the $R$-matrix (boundary) radius \cite{Tho09}.
$a_c=100$ fm is used in HHR$^\ast$.
$a_c=50$ fm \cite{Ngu12,Ngu13} is also applicable, and it would lead to a similar result.
Although $R$-matrix theory was originally developed for two-body reactions, it has been generalized for a three-body problem, straightforwardly \cite{Tho00}.

In $R$-matrix expansion (e.g.~\cite{Tho09,Des}), linearly-independent scattering waves in the interior region are expanded by
\begin{eqnarray}
  \chi^{l\hspace{1mm}{\rm in}}_{\alpha\beta}(k,\rho) &=& \mathop{\sum}_i A_{i \beta}(k) \chi^l_{\alpha i}(\rho),
  \label{eq:start}
\end{eqnarray}
where $\alpha$ and $\beta$ are the channel labels.
$k$ are hypermomenta, $k=\sqrt{2m_{\rm N} E/\hbar^2}$.
$A_{i \beta}(k)$ are the coefficients of expansion,
\begin{eqnarray}
  A_{i \beta}(k) &=&
  \frac{\hbar^2}{2m_{\rm N}} \frac{1}{\tilde{E}(l^\pi_i)-E} \mathop{\sum}_\gamma \chi^l_{\gamma i}(a_c)
  \Big[\,H^{-\hspace{1mm}\prime}_{K+3/2}(\eta_\gamma; ka_c) \delta_{\gamma\beta}
    \nonumber \\
    &-&
    S^l_{\gamma\beta}(E,a_c) H^{+\hspace{1mm}\prime}_{K+3/2}(\eta_\gamma; ka_c)\,\Big].
\end{eqnarray}
$\tilde{E}(l^\pi_i)$ and $\chi^l_{\alpha i}$ are the eigen-energies and eigenfunctions of $\hat{H}_{3\alpha}$, given in (\ref{eq:eigen-en}) and (\ref{eq:eigen-fun}).
$H^{\pm}_{K+3/2}$ are the incoming ($-$) and outgoing ($+$) Coulomb wavefunctions, defined by
\begin{eqnarray}
  H^{\pm}_{K+3/2}(\eta_\gamma; k\rho) &=& \tilde{G}_{K+3/2}(\eta_\gamma; k\rho) \pm i \tilde{F}_{K+3/2}(\eta_\gamma; k\rho),
\end{eqnarray}
where $\tilde{F}_{K+3/2}$ and $\tilde{G}_{K+3/2}$ are the regular and irregular Coulomb wavefunctions, respectively.
$S^l_{\gamma\beta}(E,a_c)$ is the three-body $S$-matrix at $\rho=a_c$,
\begin{eqnarray}
  S^l_{\alpha\beta}(E,a_c) &=& [{\cal Z}^\ast_{\alpha\beta}(E,a_c)]^{-1} {\cal Z}_{\alpha\beta}(E,a_c),
  \\
  {\cal Z}_{\alpha\beta}(E,a_c) &=& H^{-}_{K+3/2}(\eta_\alpha; ka_c) \delta_{\alpha\beta} -a_c R^l_{\alpha\beta}(E,a_c) H^{-\hspace{1mm}\prime}_{K+3/2}(\eta_\beta; ka_c).
\end{eqnarray}
$R^l_{\alpha\beta}(E,a_c)$ is the $R$-matrix,
\begin{eqnarray}
  R^l_{\alpha\beta}(E,a_c) &=& \mathop{\sum}_i \frac{\tilde{\gamma}_{\alpha i}\tilde{\gamma}_{\beta i}}{\tilde{E}(l^\pi_i)-E}.
  \label{eq:rmat}
\end{eqnarray}
$\tilde{\gamma}_{\alpha i}$ are the three-body reduced width amplitudes,
\begin{eqnarray}
  \tilde{\gamma}_{\alpha i} &=& \sqrt{\frac{\hbar^2}{2m_{\rm N} a_c}} \chi^l_{\alpha i}(a_c).
  \label{eq:rwidth}
\end{eqnarray}
Although the scattering waves are expanded as (\ref{eq:start}), additional distortion is brought in by Coulomb couplings in the external region.
To take it into consideration, CC equations are numerically solved for the external region.
In this calculation, the values of $\chi_{\alpha\beta}^{l\hspace{1mm}{\rm in}}( k,\rho)$ are used as the initial condition.
The $n_\gamma$ independent initial conditions for (\ref{eq:hscc}) make the $n_\gamma$ linearly-independent solutions.
Consequently, the wavefunctions compose the ($n_\gamma \times n_\gamma$) matrix, expressed as $\chi^{l\hspace{1mm}{\rm ext}}_{\alpha\beta}(k,\rho)$.
In \sref{sec:2.5}, I describe how to obtain $\chi^{l\hspace{1mm}{\rm ext}}_{\alpha\beta}(k,\rho)$ from $\chi^{l\hspace{1mm}{\rm in}}_{\alpha\beta}(k,\rho)$.

In accordance with the traditional procedure of solving CC equations \cite{Sat}, the numerical solutions satisfying the scattering-boundary condition are obtained from a linear-combination of $\chi^{l\hspace{1mm}{\rm ext}}_{\alpha\beta} (k,\rho)$,
\begin{eqnarray}
  \tilde{\chi}^{l}_{\gamma\gamma_0}(k,\rho) &=& \mathop{\sum}_{\gamma^\prime} C_{\gamma^\prime\gamma_0}(k) \chi^{l\hspace{1mm}{\rm ext}}_{\gamma\gamma^\prime}(k,\rho),
  \label{eq:scat}
\end{eqnarray}
where $C_{\gamma^\prime\gamma_0}$ are the coefficients of expansion.
$C_{\gamma^\prime\gamma_0}$ are determined from matching the asymptotic wavefunctions \cite{Kat23,Sat},
\begin{eqnarray}
  \tilde{\chi}^l_{\gamma\gamma_0}(k,\rho)
  \!\!&\rightarrow&\!\! \frac{i}{2}
  \Big[\,I^{(\gamma_0)}_{\gamma, K+3/2}(\eta_\gamma; k\rho_{\rm m}) -\mathop{\sum}_{\gamma^\prime} S^l_{\gamma^\prime \gamma_0}(E)
    O^{(\gamma^\prime)}_{\gamma, K+3/2}(\eta_\gamma; k\rho_{\rm m})\Big],
  \label{eq:asym-s}
\end{eqnarray}
where $\rho_{\rm m}$ denotes a matching radius.
$S^l_{\gamma^\prime \gamma_0}(E)$ is the $S$-matrix.
$I_\gamma^{(\gamma_0)}$ and $O_\gamma^{(\gamma^\prime)}$ are the incoming and outgoing coupled Coulomb wavefunctions, $I = O^\ast$,
\begin{eqnarray}
  O^{(\gamma^\prime)}_{\gamma, K+3/2}(\eta_\gamma; k\rho) &=& a_\gamma^{(\gamma^\prime)}(k,\rho) H^{+}_{\gamma, K+3/2}(\eta_\gamma; k\rho).
\end{eqnarray}
In the asymptotic region, $a_\gamma^{(\gamma^\prime)}$ behaves as $a_\gamma^{(\gamma^\prime)} \rightarrow \delta_{\gamma\gamma^\prime}$.
However, the wavefunctions even at $\rho_{\rm m}$ appear to be distorted by Coulomb couplings.
Thus, some techniques are required to cope with the long-range behavior of couplings.

The screening technique \cite{Ngu12,Ngu13} is one of them.
In this method, strength of $V^l_{\gamma\gamma^\prime}$ ($\gamma \ne \gamma^\prime$) is reduced smoothly by a factor,
\begin{eqnarray}
  V^l_{\gamma\gamma^\prime}(\rho) &\rightarrow& V^l_{\gamma\gamma^\prime}(\rho)/\{1+\exp[(\rho-\rho_{\rm sc})/a_{\rm sc}]\},
  \label{eq:scrn}
\end{eqnarray}
where $\rho_{\rm sc}$ and $a_{\rm sc}$ are the screening radius and diffuseness, $\rho_{\rm sc}=650$ fm and $a_{\rm sc}=10$ fm.
Thus, the interaction matrix is diagonalized at $\rho_{\rm m} > \rho_{\rm sc}$, and $a_\gamma^{(\gamma^\prime)}=\delta_{\gamma\gamma^\prime}$ can be used.
The screening technique may be applicable for the triple-$\alpha$ reaction because the non-diagonal part of coupling potentials is one order of magnitude weaker than the diagonal part.
In fact, $a_\gamma^{(\gamma^\prime)}=\delta_{\gamma\gamma^\prime}$ could be used without introducing the screening potential because of $[\bm{V}^l(\rho_{\rm m})-E\bm{1}]$, as demonstrated in \sref{sec:5.1}.
In the comparison, a ratio of the calculations between screening and no-screening is defined as
\begin{eqnarray}
  d_{\rm scrn} = \left|\frac{\sigma_{\rm g} - \sigma_{\rm g}^{\rm NS}}{\sigma_{\rm g}^{\rm NS}} \right| \times 100,
  \label{eq:diff_scrn}
\end{eqnarray}
where $\sigma_{\rm g}$ and $\sigma_{\rm g}^{\rm NS}$ represent the calculated cross sections with and without the screening potential.

Using (\ref{eq:start}) and $C_{\gamma^\prime \gamma_0}$, I obtain the internal scattering wavefunctions satisfying the asymptotic boundary condition as
\begin{eqnarray}
  \tilde{\chi}^{l\hspace{1mm} {\rm in}}_{\gamma\gamma_0}(k,\rho)
  &=& \mathop{\sum}_{\gamma^\prime} C_{\gamma^\prime\gamma_0}(k)
  \chi^{l\hspace{1mm}{\rm in}}_{\gamma\gamma^\prime}(k,\rho) \nonumber \\
  &=&
  \mathop{\sum}_i D_{i\gamma_0}(k) \chi^l_{\gamma i} (\rho), \\
  D_{i\gamma_0}(k) &\equiv& \mathop{\sum}_{\gamma^\prime} C_{\gamma^\prime\gamma_0}(k) A_{i\gamma^\prime}(k).
\end{eqnarray}
The three-body continuum states are therefore expanded as
\begin{eqnarray}
  \Psi_{lm}^{(+)} &=& \frac{1}{(k\rho)^{5/2}}\mathop{\sum}_{n \gamma\gamma_0} i^K
  \big[\mathop{\sum}_{i} c^n_{\gamma i} D_{i\gamma_0}(k) \big] \,\varphi_n^K(\rho)
  \Phi^{\gamma}_{lm}(\Omega_5^{(3)}) \Phi^{\gamma_0\,\ast}_{lm}(\Omega^k_5),
  \label{eq:continuum}
\end{eqnarray}
where $\Omega^k_5$ is the angle of $\bm{k}$.
In the external region, they are given with the numerical solutions of (\ref{eq:scat}) as
\begin{eqnarray}
  \Psi_{lm}^{(+)} &=& \frac{1}{(k\rho)^{5/2}}\mathop{\sum}_{\gamma\gamma_0} i^K \tilde{\chi}^{l}_{\gamma\gamma_0}(k,\rho)
  \Phi^{\gamma}_{lm}(\Omega_5^{(3)}) \Phi^{\gamma_0\,\ast}_{lm}(\Omega^k_5).
  \label{eq:continuum-ext}
\end{eqnarray}

\subsection{$R$-matrix propagation}\label{sec:2.5}

\begin{figure}[t] 
  \begin{center}
    \includegraphics[width=0.65\linewidth]{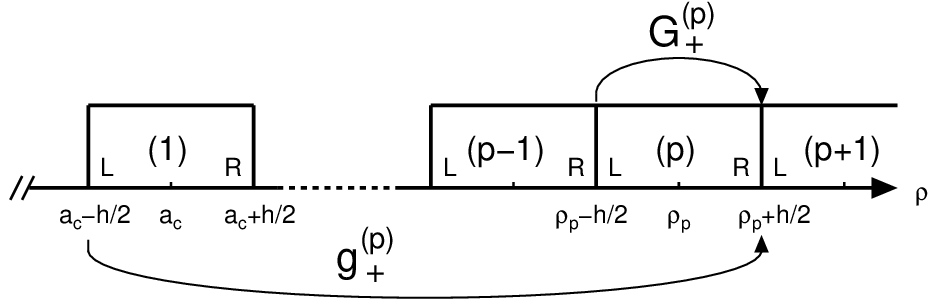}
  \end{center}
  \caption{\label{fig:sector}
Schematic diagram of sectors in $R$-matrix propagation.
  }
\end{figure}

In $R$-matrix propagation \cite{Ngu12,Ngu13}, the external region is divided into sectors, as in \fref{fig:sector}, and the $R$-matrix in the $(p-1)$ sector $\bm{R}^{(p-1)}$ is propagated to $\bm{R}^{(p)}$, as follows:
\begin{eqnarray}
  \bm{R}^{(p)} &=& \frac{1}{\rho^{(p)}_{\rm R}} \left[ \bm{G}_2^{(p)}
     \big(\,\bm{G}_4^{(p)}-\rho^{(p-1)}_{\rm R}\bm{R}^{(p-1)}\big)^{-1}
     \bm{G}_3^{(p)} -\bm{G}_1^{(p)} \right],
\end{eqnarray}
where $p\ge2$, $\rho^{(p-1)}_{\rm R}=\rho_p-h/2=\rho^{(p)}_{\rm L}$, $\rho^{(p)}_{\rm R}=\rho_p+h/2$.
$\rho_p$ is the center of the $p$ sector.
$h$ is a sector size, and it is set to $h=0.01$ fm in HHR$^\ast$.
The subscript R (L) means the right (left) boundary of the sector.
$\bm{G}_i^{(p)}$ ($i=1,2,3,4$) are the propagating matrices, defined by
\begin{eqnarray}
  \bm{G}^{(p)}_i &=& {\cal T}^{(p)} \bm{\Lambda}^{(p)}_i {}^t{\cal T}^{(p)}.
\end{eqnarray}
${\cal T}^{(p)}$ is a unitary matrix for diagonalization of a matrix $\bm{Q}^{(p)}$,
\begin{eqnarray}
  {}^t{\cal T}^{(p)}\bm{Q}^{(p)}{\cal T}^{(p)}
  &=& \left(\begin{array}{cccccc}
    \lambda_1^2(p) &     0      & \ldots  & 0  \\
        0      & \lambda_2^2(p) & \ldots  & 0  \\
      \vdots   &  \vdots    & \ddots & \vdots \\
        0      &     0      & \ldots  & \lambda_{n_\gamma}^2(p) \\
            \end{array}\right),
\end{eqnarray}
\begin{eqnarray}
  Q^{(p)}_{\gamma^\prime \gamma} &=& \left[\frac{(K+3/2)(K+5/2)}{\rho_p^2}+\epsilon\,\right] \delta_{\gamma^\prime\gamma}
  -U^l_{\gamma^\prime\gamma}(\rho_p).
\end{eqnarray}
$\bm{\Lambda}^{(p)}_i$ is defined as a diagonal matrix:
\numparts
\begin{eqnarray}
  (\Lambda_1^{(p)})_{\gamma^\prime\gamma} = (\Lambda_4^{(p)})_{\gamma^\prime\gamma} &=& \delta_{\gamma^\prime\gamma} \left\{\begin{array}{ll}
  -\frac{\displaystyle{1}}{\displaystyle{|\lambda_\gamma(p)| \tanh |h\lambda_\gamma(p)|}} & \lambda^2_{\gamma}(p) > 0\\
  & \\
  \frac{\displaystyle{1}}{\displaystyle{|\lambda_\gamma(p)| \tan  |h\lambda_\gamma(p)|}} & \lambda^2_{\gamma}(p) \le 0\\
  \end{array}\right.\\
  && \nonumber\\
  (\Lambda_2^{(p)})_{\gamma^\prime\gamma} = (\Lambda_3^{(p)})_{\gamma^\prime\gamma} &=& \delta_{\gamma^\prime\gamma} \left\{\begin{array}{ll}
  -\frac{\displaystyle{1}}{\displaystyle{|\lambda_\gamma(p)| \sinh |h\lambda_\gamma(p)|}} & \lambda^2_{\gamma}(p) > 0\\
  & \\
  \frac{\displaystyle{1}}{\displaystyle{|\lambda_\gamma(p)| \sin  |h\lambda_\gamma(p)|}} & \lambda^2_{\gamma}(p) \le 0\\
  \end{array}\right.
\end{eqnarray}
\endnumparts

The corresponding wavefunctions and their derivatives at the left boundary of the $p$ sector are calculated from those of the $(p-1)$ sector, as follows:
\begin{eqnarray}
  \left(\begin{array}{c}
    \!\! \bm{\chi}^{(p)}_{\rm L} \!\! \\
    \\
    \!\! \bm{\chi}^{(p)\,\prime }_{\rm L} \!\! \\
  \end{array}\right)
  &=&
  \bm{G}^{(p-1)}_+
  \left(\begin{array}{c}
    \!\! \bm{\chi}^{(p-1)}_{\rm L} \!\!\!\!  \\
    \\
    \!\! \bm{\chi}^{(p-1)\,\prime}_{\rm L} \!\!\!\!  \\
  \end{array}\right),
  \label{eq:localwf}
\end{eqnarray}
\begin{eqnarray}
  \bm{G}^{(p)}_+ \!&\equiv&\!
  \left(\begin{array}{cc}
    \!\! \bm{G}^{(p)}_1(\bm{G}^{(p)}_3)^{-1}  \,\,& \,\,
    \left[ \bm{G}^{(p)}_2 \!\! -\! \bm{G}^{(p)}_1(\bm{G}^{(p)}_3)^{-1}\bm{G}^{(p)}_4\right] \!\! \\
    \\
    -(\bm{G}^{(p)}_3)^{-1} &
    (\bm{G}^{(p)}_3)^{-1}\bm{G}^{(p)}_4 \\
  \end{array}\!\right).
\end{eqnarray}
Therefore, $\chi^{l\hspace{1mm}{\rm ext}}_{\alpha\beta}(k,\rho)$ and $d\chi^{l\hspace{1mm}{\rm ext}}_{\alpha\beta}(k,\rho)/d\rho$ are obtained recursively from $\rho=a_c-h/2$.
The initial values are given by (\ref{eq:eigen-fun}) and (\ref{eq:start}) with (\ref{eq:gwf0}) and (\ref{eq:dgwf0}).
To execute stable calculations, quadruple precision is required.

The global propagation matrices $\bm{g}^{(p)}_i$ \cite{Ngu12,Ngu13} are calculated recursively from $\bm{G}^{(p)}_i$, as follows:
\numparts
\begin{eqnarray}
  \bm{g}_1^{(p)} &=& \bm{G}_1^{(p)} -\bm{G}_2^{(p)} (\,\bm{g}_1^{(p-1)}+\bm{G}_4^{(p)}\,)^{-1} \bm{G}_3^{(p)}, \\
  \bm{g}_2^{(p)} &=& \bm{G}_2^{(p)}  (\,\bm{g}_1^{(p-1)}+\bm{G}_4^{(p)}\,)^{-1} \bm{g}_2^{(p-1)}, \\
  \bm{g}_3^{(p)} &=& \bm{g}_3^{(p-1)}(\,\bm{g}_1^{(p-1)}+\bm{G}_4^{(p)}\,)^{-1} \bm{G}_3^{(p)}, \\
  \bm{g}_4^{(p)} &=& \bm{g}_4^{(p-1)} -\bm{g}_3^{(p-1)}(\,\bm{g}_1^{(p-1)}+\bm{G}_4^{(p)}\,)^{-1}\bm{g}_2^{(p-1)},
\end{eqnarray}
\endnumparts
where $p\ge2$.
The initial values ($p=1$) are $\bm{g}_1^{(1)}=\bm{G}_1^{(1)}$, $\bm{g}_2^{(1)}=\bm{G}_2^{(1)}$, $\bm{g}_3^{(1)}=\bm{G}_3^{(1)}$ and $\bm{g}_4^{(1)}=\bm{G}_4^{(1)}$.
The global forward propagation matrix is defined by
\begin{eqnarray}
  \bm{g}^{(p)}_+
  &\equiv&
  \left(\begin{array}{cc}
    \bm{g}_1^{(p)}(\bm{g}_3^{(p)})^{-1} & \left[\bm{g}_2^{(p)}-\bm{g}_1^{(p)}(\bm{g}_3^{(p)})^{-1}\bm{g}_4^{(p)}\right] \\
    &\\
    -(\bm{g}_3^{(p)})^{-1} & (\bm{g}_3^{(p)})^{-1}\bm{g}_4^{(p)}  \\
  \end{array}\right).
\end{eqnarray}
The wavefunctions and their derivatives at the right boundary of the $p$th sector are obtained from those at the left boundary of the first sector, as follows:
\begin{eqnarray}
  \left(\begin{array}{c}
    \bm{\chi}_{\rm R}^{(p)} \\
    \\
    \bm{\chi}_{\rm R}^{(p)\,\prime} \\
  \end{array}\right)
  &=&
  \bm{g}^{(p)}_+
  \left(\begin{array}{c}
    \bm{\chi}_{\rm L}^{(1)} \\
    \\
    \bm{\chi}_{\rm L}^{(1)\,\prime}\\
  \end{array}\right).
\end{eqnarray}
$\bm{g}^{(p)}_+$ is independent of the short-range 3$\alpha$ potentials $V^l_{3\alpha}(\rho)$, and it includes all the effects of the long-range Coulomb couplings.
If $\rho_{\rm m}= \rho_p+h/2$, the wavefunctions and derivatives at $\rho_{\rm m}$ are given by the single operation at $a_c-h/2$,
\numparts
\begin{eqnarray}
  \chi^{l\hspace{1mm}{\rm ext}}_{\alpha\beta}(k,\rho_{\rm m})&=&
  \mathop{\sum}_\gamma \left[\bm{g}_1^{(p)}(\bm{g}_3^{(p)})^{-1} \right]_{\alpha\gamma}\chi^{l\hspace{1mm}{\rm in}}_{\gamma\beta}(k,a_c-h/2)
  \nonumber \\
  &+&
  \mathop{\sum}_\gamma \left[\bm{g}_2^{(p)}-\bm{g}_1^{(p)}(\bm{g}_3^{(p)})^{-1}\bm{g}_4^{(p)}\right]_{\alpha\gamma}\frac{d\chi^{l\hspace{1mm}{\rm in}}_{\gamma\beta}(k,a_c-h/2)}{d\rho},
  \label{eq:chi_gp}
  \\
  \frac{d\chi^{l\hspace{1mm}{\rm ext}}_{\alpha\beta}(k,\rho_{\rm m})}{d\rho}&=&
  \mathop{\sum}_\gamma \left[-(\bm{g}_3^{(p)})^{-1}\right]_{\alpha\gamma}\chi^{l\hspace{1mm}{\rm in}}_{\gamma\beta}(k,a_c-h/2)
  \nonumber \\
  &+&
  \mathop{\sum}_\gamma \left[(\bm{g}_3^{(p)})^{-1}\bm{g}_4^{(p)}\right]_{\alpha\gamma}\frac{d\chi^{l\hspace{1mm}{\rm in}}_{\gamma\beta}(k,a_c-h/2)}{d\rho}.
  \label{eq:dchi_gp}
\end{eqnarray}
\endnumparts
These are the same as those obtained locally from (\ref{eq:localwf}).
Therefore, the values calculated from (\ref{eq:chi_gp}) and (\ref{eq:dchi_gp}) or (\ref{eq:localwf}) are used for the asymptotic matching of (\ref{eq:scat}) and (\ref{eq:asym-s}).
The matching radius is $\rho_{\rm m}= (3000+0.005)$ fm \cite{Ngu12,Ngu13}.
If the screening factor of (\ref{eq:scrn}) is switched on, $\rho_{\rm m}= (1500+0.005)$ fm is also applicable because the coupling potentials are approximately diagonal at $\rho_{\rm m}$.

Likewise, the global backward propagation matrix is defined by
\begin{eqnarray}
  \bm{g}^{(p)}_-
  &\equiv&
  (\bm{g}^{(p)}_+)^{-1} \hspace{1mm}=\hspace{1mm}
  \left(\begin{array}{cc}
    \bm{g}_4^{(p)}(\bm{g}_2^{(p)})^{-1} & \left[\bm{g}_4^{(p)}(\bm{g}_2^{(p)})^{-1}\bm{g}_1^{(p)}-\bm{g}_3^{(p)}\right]\\
    &\\
    (\bm{g}_2^{(p)})^{-1} & (\bm{g}_2^{(p)})^{-1}\bm{g}_1^{(p)}  \\
  \end{array}\right),
  \label{eq:bwpropmat}
\end{eqnarray}
\begin{eqnarray}
  \left(\begin{array}{c}
    \bm{\chi}_{\rm L}^{(1)} \\
    \\
    \bm{\chi}_{\rm L}^{(1)\,\prime}\\
  \end{array}\right)
  &=&
  \bm{g}^{(p)}_-
  \left(\begin{array}{c}
    \bm{\chi}_{\rm R}^{(p)} \\
    \\
    \bm{\chi}_{\rm R}^{(p)\,\prime} \\
  \end{array}\right).
  \label{eq:bwprop}
\end{eqnarray}
However, $\bm{g}^{(p)}_-$ may be devastated by numerical processing at the very low energies.
Due to this breakdown, equation~(\ref{eq:bwprop}) is not always true in numerical calculations.
This global backward propagation is not used in HHR$^\ast$.

\section{Formulae for the triple-$\alpha$ reaction rates}\label{sec:3}

\subsection{Transition probability and the reaction rates}\label{sec:3.1}

The reduced transition strength \cite{Ngu12,Ngu13,Tho09} from the 0$^+$ continuum states to the 2$^+_1$ bound state in $^{12}$C is calculated from
\begin{eqnarray}
  \frac{dB(E2; l\rightarrow l^\prime)}{dE}
  \!\!&=&\!\!
  \frac{1}{(2l+1)}
  \frac{2k^5}{\pi\hbar v}\mathop{\sum}_{m^\prime m q} \!
  \int \Big|\langle \Psi_{f,l^\prime m^\prime}\,|\,{\cal M}^{\rm E}_{2q}\,|\,\Psi_{lm}^{(+)}\rangle\Big|^2 \!d\Omega^k_5,
\end{eqnarray}
where $l=0$, $l^\prime=2$ and $f=1$.
$q$ is the projection of multipolarity.
${\cal M}^{\rm E}_{2q}$ denotes the electric quadrupole ($E$2) operator,
\begin{eqnarray}
  {\cal M}^{\rm E}_{2q} &=&
  2e\big[ r_1^2 Y_{2q}(\hat{\bm{r}}_1) +r_2^2 Y_{2q}(\hat{\bm{r}}_2) +r_3^2 Y_{2q}(\hat{\bm{r}}_3) \big]
  \nonumber \\
  &=&
  \frac{e}{2} \,\rho^2\,\big[ \sin^2\theta_3 Y_{2q}(\hat{\bm{x}})+\cos^2\theta_3 Y_{2q}(\hat{\bm{y}})\big],
  \label{eq:me2q}
\end{eqnarray}
where $\bm{r}_i$ is a distance from the center of mass.
Replacing $\Psi_{f,l^\prime m^\prime}$, $\Psi_{lm}^{(+)}$ and ${\cal M}^{\rm E}_{2q}$ with (\ref{eq:eigen-fun-tot}), (\ref{eq:continuum}), (\ref{eq:continuum-ext}) and (\ref{eq:me2q}), $dB(E2)/dE$ is deduced, as follows:
\begin{eqnarray}
  \frac{dB(E2; 0^+ \rightarrow 2^+_1)}{dE}
  \!&=&\!
  \frac{5e^2}{2\pi\hbar v} \mathop{\sum}_{\gamma_0}
  \Big| \mathop{\sum}_{\gamma^\prime\gamma} M^{2^+_1 0^+}_{\gamma^\prime\gamma\gamma_0}(k)
  \Big( f^x_{\gamma^\prime\gamma} \zeta_x + f^y_{\gamma^\prime\gamma} \zeta_y \Big)\,\Big|^2,
  \label{eq:be2}
\end{eqnarray}
where
\begin{eqnarray}
  M^{2^+_1\,0^+}_{\gamma^\prime\gamma\gamma_0}(k)
  = M^{2^+_1\,0^+\,{\rm Int}}_{\gamma^\prime\gamma\gamma_0}(k) + M^{2^+_1\,0^+\,{\rm Ext}}_{\gamma^\prime\gamma\gamma_0}(k),
  \label{eq:Mgg}
\end{eqnarray}
\numparts
\begin{eqnarray}
  f^x_{\gamma^\prime\gamma} = 
  \int \phi^{l^\prime_x l^\prime_y}_{K^\prime} (\theta_3) \sin^4\theta_3 \cos^2\theta_3 \phi^{l_x l_y}_K (\theta_3) d\theta_3,
  \\
  f^y_{\gamma^\prime\gamma} = 
  \int \phi^{l^\prime_x l^\prime_y}_{K^\prime} (\theta_3) \sin^2\theta_3 \cos^4\theta_3 \phi^{l_x l_y}_K (\theta_3) d\theta_3,
  \\
  \zeta_x = \delta_{l_y l_y^\prime} \delta_{l_x l_y} \frac{(-)^{(l_x^\prime+l_x)/2}}{\sqrt{4\pi}} \hat{l}_x^\prime
  \left(\begin{array}{ccc}
    l^\prime_x & 2 & l_x \\
    0       & 0 & 0 \\
  \end{array}\right),
  \\
  \zeta_y = \delta_{l_x l_x^\prime} \delta_{l_x l_y} \frac{(-)^{(l_y^\prime+l_y)/2}}{\sqrt{4\pi}} \hat{l}_y^\prime
  \left(\begin{array}{ccc}
    l^\prime_y & 2 & l_y \\
    0       & 0 & 0 \\
  \end{array}\right).
\end{eqnarray}
\endnumparts
$M^{2^+_1\,0^+\,{\rm Int}}_{\gamma^\prime\gamma\gamma_0}(k)$ and $M^{2^+_1\,0^+\,{\rm Ext}}_{\gamma^\prime\gamma\gamma_0}(k)$ are the interior and external components of transition amplitudes,
\numparts
\begin{eqnarray}
  M^{2^+_1\,0^+\,{\rm Int}}_{\gamma^\prime\gamma\gamma_0}(k) &=&
  i^K \mathop{\sum}_i D_{i \gamma_0}(k) \int_0^{a_c} \chi^{2^+}_{\gamma^\prime 1}(\rho) \chi^{0^+}_{\gamma i}(\rho) \rho^2 d\rho,
  \label{eq:Mgg-in}
  \\
  M^{2^+_1\,0^+\,{\rm Ext}}_{\gamma^\prime\gamma\gamma_0}(k) &=&
  i^K \int_{a_c}^\infty \chi^{2^+}_{\gamma^\prime 1}(\rho) \tilde{\chi}^{0^+}_{\gamma \gamma_0}(k,\rho) \rho^2 d\rho.
  \label{eq:Mgg-ext}
\end{eqnarray}
\endnumparts
The integrals of (\ref{eq:Mgg-ext}) converge by $\rho=200$ fm.
The reduced transition probability $B(E2;0^+_2\rightarrow 2^+_1)$ is obtained from the integral of (\ref{eq:be2}) around $E$(0$^+_2$), and the $\gamma$-decay width of 0$^+_2$ is given by
\begin{eqnarray}
  \Gamma_\gamma(0^+_2) &=& \frac{4\pi}{75}\Big( \frac{\bar{E}_{\rm g}}{\hbar c}\Big)^5 \int_{E(0^+_2)-\Delta\epsilon}^{E(0^+_2)+\Delta\epsilon} \frac{dB(E2;0^+_2\rightarrow 2^+_1)}{dE} dE,
  \label{eq:gg}
\end{eqnarray}
where $\bar{E}_{\rm g}$ is $\gamma$-ray energy, $\bar{E}_{\rm g}=E(0^+_2)-E(2^+_1)$.
The reduced transition probability from 2$^+_1$ to the ground state is 
\begin{eqnarray}
  B(E2;2^+_1\rightarrow 0^+_1) 
  = \frac{e^2}{4} \Big| \mathop{\sum}_{\gamma^\prime\gamma} M^{2^+_1\,0^+_1}_{\gamma^\prime\gamma}
  \Big( f^x_{\gamma^\prime\gamma} \zeta_x + f^y_{\gamma^\prime\gamma} \zeta_y \Big)\,\Big|^2,
  \label{eq:be2_2gs}
  \\
  M^{2^+_1\,0^+_1}_{\gamma^\prime\gamma} =
  \int_0^{a_c} \chi^{2^+}_{\gamma^\prime 1}(\rho) \chi^{0^+}_{\gamma 1}(\rho)\rho^2 d\rho.
\end{eqnarray}

The photo-disintegration of $^{12}$C($2^+_1 \rightarrow 0^+$) is the inverse reaction of the triple-$\alpha$ process, and its cross sections are given in \cite{Ngu12,Ngu13,Tho09} as
\begin{eqnarray}
  \sigma_{\rm g}(E)
  &=& \frac{4\pi^3}{75}\left(\frac{E_{\rm g}}{\hbar c}\right)^3 \frac{1}{5}\,\frac{dB(E2;0^+\rightarrow 2^+_1)}{dE},
  \label{eq:sig}
\end{eqnarray}
where $E_{\rm g}=E-E(2^+_1)$.
The reaction rates of 3$\alpha \rightarrow ^{12}$C(2$^+_1$) + $\gamma$ are 
\begin{eqnarray}
  R_{3\alpha}(E) &=&
  N_{\rm A}^2\frac{480\pi}{(\mu_{12}\mu_{(12)3})^{3/2}} \frac{\hbar^3}{c^2}\, \frac{E^2_{\rm g}}{E^2} \,\sigma_{\rm g}(E),
  \label{eq:r3a}
\end{eqnarray}
where $N_{\rm A}$ is the Avogadro constant.
The energy-averaged reaction rates over the three-body Maxwell-Boltzmann distribution are given by
\begin{eqnarray}
  \langle R_{3\alpha}\rangle &=&
  \frac{1}{2(k_{\rm B} T)^3} \int^\infty_0 R_{3\alpha}(E) E^2 \exp\Big(-\frac{E}{k_{\rm B} T}\Big) dE,
  \label{eq:rates3a}
\end{eqnarray}
where $k_{\rm B}$ and $T$ are the Boltzmann constant and temperature, respectively.

\subsection{Analytic expression of the reaction rates}\label{sec:3.2}

If a three-body type of $S$-factors is defined as
\begin{eqnarray}
  S_{3\alpha}(E) &\equiv&
  E_{\rm g}^2 \sigma_{\rm g}(E) \exp\left(\frac{2\pi\eta_0}{\sqrt{E}} +a E\right)
  \nonumber \\
  &=& s_0 \,\big(1 + s_1 E + s_2 E^2\big),
  \label{eq:3a_fact}
\end{eqnarray}
the integral of energy distribution in (\ref{eq:rates3a}) is performed with
\begin{eqnarray}
\fl  E_{\rm g}^2 \sigma_{\rm g}(E) \exp\Big(-\frac{E}{k_{\rm B}T}\Big)
  &=&
  s_0\,\big(1 + s_1 E + s_2 E^2\big)\exp\Big[ -\frac{(1+a k_{\rm B} T)E}{k_{\rm B} T} -\frac{2\pi\eta_0}{\sqrt{E}} \,\Big]
  \nonumber\\
  &=&
  S_{3\alpha}(E) \exp\Big[ -\left(\frac{E-E_0}{\Delta/2}\right)^2 -\frac{3E_0}{T_a} -\bar{g} \,\Big],
\end{eqnarray}
where $T_a = (k_{\rm B} T)/(1 + a k_{\rm B} T)$.
$\bar{g}$ denotes non-Gaussian components.
$E_0$ and $\Delta$ are the Gamow peak energy and energy-window for the direct triple-$\alpha$ process,
\begin{eqnarray}
  E_0&=&(\pi \eta_0 T_a)^{2/3},
  \label{eq:GE}\\
  \Delta&=&(4/\sqrt{3})\,(\pi\eta_0)^{1/3}T_a^{5/6}.
  \label{eq:GW}
\end{eqnarray}
$s_0$, $s_1$, $s_2$, $\eta_0$ and $a$ are determined from the calculated $E_{\rm g}^2\sigma_{\rm g}(E)$.
With this expansion, the non-resonant reaction rates are analytically expressed as
\begin{eqnarray}
  \langle R_{3\alpha}\rangle_{\rm NR}
  \!&=&\!
  N_{\rm A}^2\,\frac{45\pi\sqrt{\pi}}{2m_{\rm N}^3}\frac{\hbar^3}{c^2}
  \,\frac{(\pi\eta_0)^{1/3}}{(k_{\rm B} T)^{13/6}(1+a k_{\rm B} T)^{5/6}}
  \,\tilde{S}(T_a)
  \exp\Big[ -\frac{3(\pi\eta_0)^{2/3}}{T_a^{1/3}} \Big],
  \label{eq:rates_nr}
  \nonumber\\
\end{eqnarray}
\begin{eqnarray}
  \tilde{S}(T_a)
  \!&\approx&\!
     s_0 \,\Big[\,     \left(1+\frac{5T_a}{36E_0}\right)
    +s_1 E_0  \left(1+\frac{35T_a}{36E_0}\right)
    +s_2 E_0^2\left(1+\frac{89T_a}{36E_0}\right)\,\Big].
  \label{eq:rates_st}
  \nonumber \\
\end{eqnarray}
The significant term at low temperatures may be represented by
\begin{eqnarray}
  \langle R_{3\alpha}\rangle_{\rm NR} &\approx& \frac{\tilde{b}_0}{T_9^{13/6}} \exp\Big(-\frac{\tilde{b}_1}{T_9^{1/3}}\Big).
  \label{eq:ex-rates_2}
\end{eqnarray}
This simplified version is used for REACLIB translation.
$\tilde{b}_0$ and $\tilde{b}_1$ are adjusted so as to minimize a difference from the expression of (\ref{eq:rates_nr}).

The resonant contribution in (\ref{eq:rates3a}) is analytically expressed as
\begin{eqnarray}
  \langle R_{3\alpha}\rangle_{\rm R}
  &=&
  \frac{9\sqrt{3}\pi^3}{4} \frac{N_{\rm A}^2 \,\hbar^5}{m_{\rm N}^3}
  \frac{\Gamma_\gamma(0^+_2)}{(k_{\rm B} T)^3} \,\exp\Big(-\frac{E(0^+_2)}{k_{\rm B} T}\Big) \nonumber \\
  &\approx& 7.605\times10^{-9}\frac{\Gamma_\gamma(0^+_2)}{T_9^3}\exp\Big(-\frac{11.605\,E(0^+_2)}{T_9}\Big),
  \label{eq:rates_r}
\end{eqnarray}
where $\Gamma_\gamma(0^+_2)$ is in meV, and $E(0^+_2)$ is in MeV.
$\langle R_{3\alpha}\rangle_{\rm R}$ is given in units of cm$^6$mol$^{-2}$s$^{-1}$.

\section{Interaction potentials and the lowest three states in $^{12}$C}\label{sec:4}

In this section, I describe the interaction potentials for $\alpha$+$\alpha$ and 3$\alpha$, and I examine the generated states to discuss the available range of 3$\alpha$ potentials.

\begin{table}[t] 
    \caption{\label{tb:vaa}
Parameters of $\alpha$+$\alpha$ potentials in (\ref{eq:vaa}).
The values of AB are taken from \cite{Ngu13,Ali66,Fed96}.
CD is based on \cite{Buc77,Sat79,Bay87,Ber77}.
The values of CD are determined so as to reproduce the resonant energy and width of $^8$Be(0$^+_1$) and experimental elastic phase shifts \cite{Hey56,Nil58,Tom63,Chi74}.
The values of $J_{\rm v}$, defined by (\ref{eq:jv}), are also listed as net strength of potentials.
    }
    \begin{indented}
    \lineup
    \item[]\begin{tabular}{@{}*{6}{l}c}
      \br
      & $V_{\rm c0}$(MeV) & $V_{\rm c2}$(MeV) & $r_{\rm c}$(fm) & $V_0$(MeV) & $r_{\rm v}$(fm) & $J_{\rm v}$(MeV fm$^3$) \\
      \mr
      AB & 125.0  &\020.0 & 1.53 & \0$-30.18$ & 2.85 & 243 \\
      CD & 295.45 & 197.8 & 1.57 &  $-118.7$  & 2.16 & 416 \\
      \br
    \end{tabular}
    \end{indented}
\end{table}

\subsection{$\alpha$+$\alpha$ potentials}\label{sec:4.1}

\begin{figure}[t] 
  \begin{center}
    \includegraphics[width=0.55\linewidth]{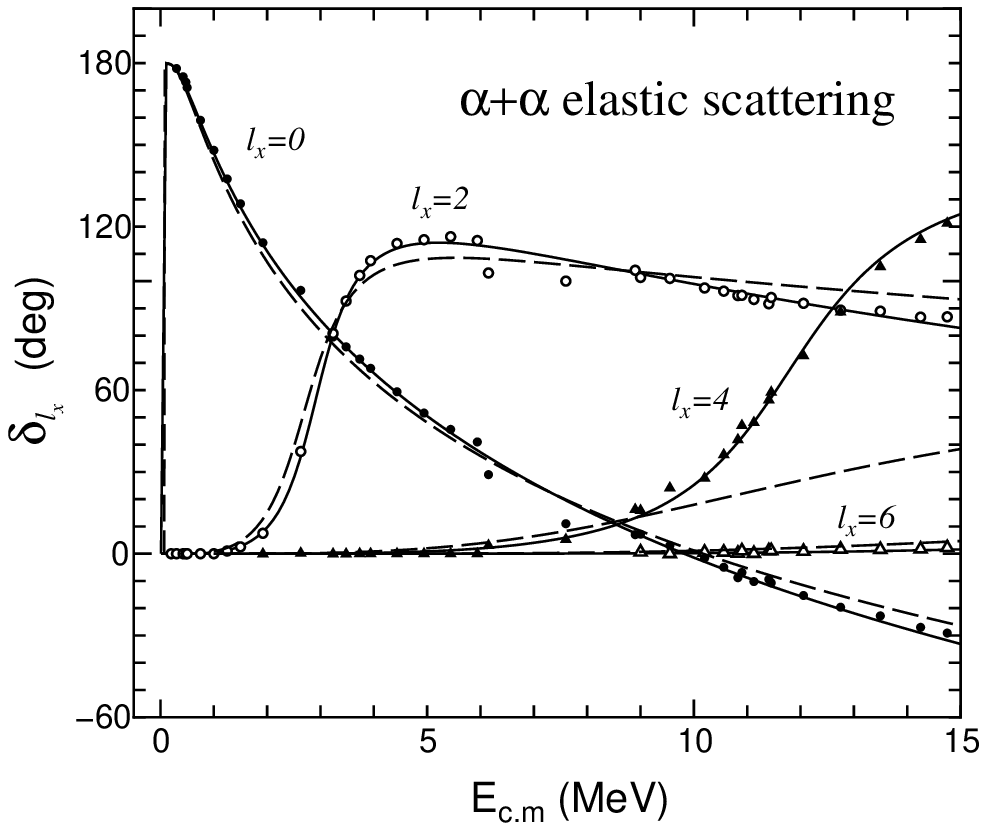}
  \end{center}
  \caption{\label{fig:ps-aa}
Phase shifts of $\alpha$+$\alpha$ elastic scattering.
The solid and dashed curves are the results calculated from the CD and AB potentials, respectively.
The experimental data are taken from \cite{Hey56,Nil58,Tom63,Chi74}.
  }
\end{figure}

Two types of interaction potentials between $\alpha$+$\alpha$ are adopted in (\ref{eq:fkk}).
One is the so-called modified Ali-Bodmer (AB) potential \cite{Ali66,Fed96,Ngu13,Sun16,Ish13}.
The other is a deep potential based on the double folding (DF) model \cite{Buc77,Sat79}.
The repulsive core potentials are introduced so as to satisfy the Pauli principle \cite{Bay87}.
This phase-equivalent potential is referred to as the core plus deep (CD) potential in the present article.
AB and CD are given in
\begin{eqnarray}
  V^{l_{x_i}}_{\rm N}(\bar{x}_i) &=& V_{{\rm c} l_{x_i}}\hat{\bf P}_{l_{x_i}} \exp\big[-(\bar{x}_i/r_{\rm c})^2\big] +V_0 \exp\big[-(\bar{x}_i/r_{\rm v})^2\big],
  \label{eq:vaa}
\end{eqnarray}
for $l_{x_i}=$ even.
$\hat{\bf P}_{l_{x_i}}$ are Pauli exclusion operators.
$V_{\rm c0}$, $V_{\rm c2}$, $V_0$, $r_{\rm c}$ and $r_{\rm v}$ are listed in \tref{tb:vaa}.
For CD, the parameters are determined so as to reproduce the resonant energy and width of $^8$Be(0$^+_1$) and the experimental phase shifts of elastic scattering \cite{Hey56,Nil58,Tom63,Chi74}.
The calculated phase shifts are compared with the experimental data in \fref{fig:ps-aa}.
The solid and dashed curves are the results obtained from the CD and AB potentials, respectively.
As shown in \fref{fig:ps-aa}, AB cannot reproduce the phase shifts of $l_x=4$, because it does not make the 4$^+$ resonance at $E_{c.m.}\approx 12$ MeV.
As a value of net strength, the volume integral of potentials per nucleon pair $J_{\rm v}$ is defined by
\begin{eqnarray}
  J_{\rm v} &=& \frac{\pi^{3/2} |V_0| r_{\rm v}^3}{16}.
  \label{eq:jv}
\end{eqnarray}
From the values of $J_{\rm v}$ in \tref{tb:vaa}, AB is found to be shallower than CD.
$J_{\rm v}$ of CD seems to be consistent with that of the studies in \cite{Sat79,Bra97,Kat13}.
AB is weak to generate correct couplings.

\begin{figure}[t] 
  \begin{center}
    \includegraphics[width=0.75\linewidth]{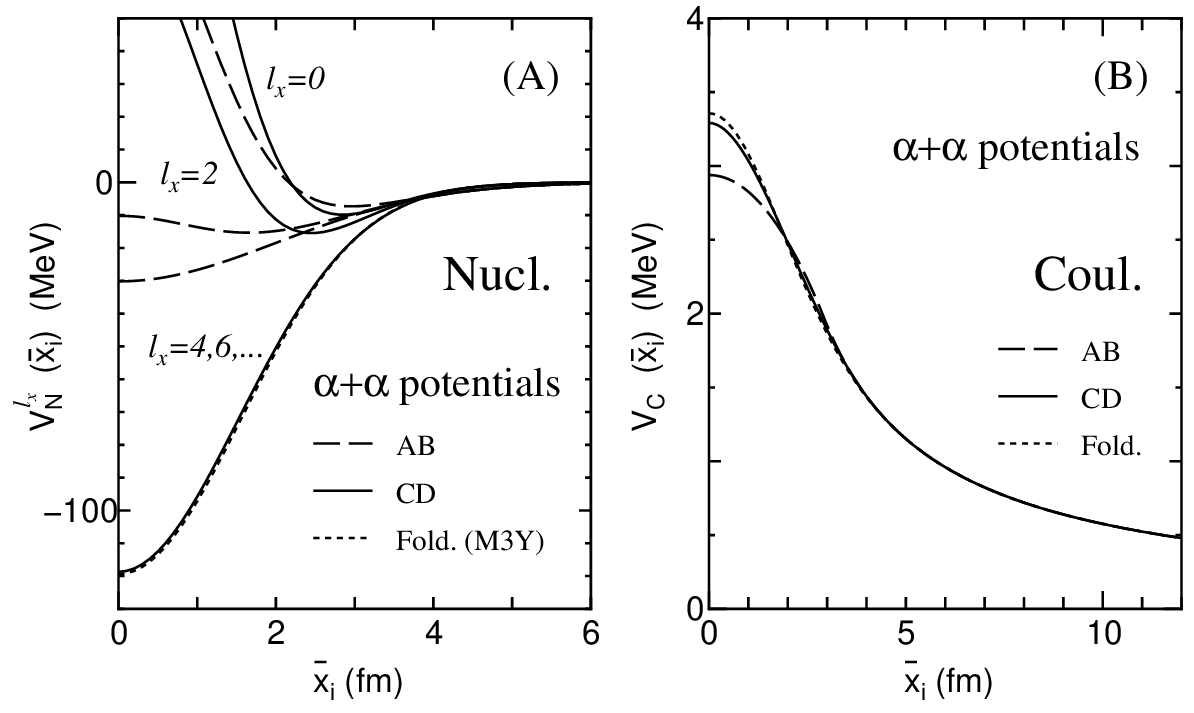}
  \end{center}
  \caption{\label{fig:vaa}
$\alpha$+$\alpha$ potentials: (A) Nuclear and (B) Coulomb parts.
The solid and dashed curves are the CD and AB potentials, respectively.
The dotted curves are calculated from the DF procedure \cite{Sat79,Bra97} with M3Y \cite{Ber77} and Coulomb interaction.
  }
\end{figure}

AB and CD are compared with the DF potential \cite{Sat79,Bra97} in \fref{fig:vaa}(A).
The dashed, solid and dotted curves are the AB, CD and DF potentials, respectively.
The DF potential is calculated from
\begin{eqnarray}
  V_{\rm F}(\bar{x}_i) &=& \int \rho_\alpha(r_j)\rho_\alpha(r_k) v_{\rm nn}(r_k+\bar{x}_i-r_j) dr_j dr_k,
  \label{eq:DFpot}
\end{eqnarray}
where $r_j$ and $r_k$ are the internal coordinates of $^4$He.
$v_{\rm nn}$ is M3Y effective nucleon-nucleon interaction \cite{Ber77} with single-nucleon exchange.
$\rho_\alpha(r)$ is $(0s)^4$ shell model density of $^4$He, given by $\rho_\alpha(r)= 0.41251 \exp[-(r/1.2031)^2]$.
The range is determined so as to reproduce the experimental charge radius, 1.6768 fm \cite{deV87}.
The rms radius of this density is $\langle r^2_\alpha \rangle^{1/2} = 1.4735$ fm, and it is also used in (\ref{eq:rms}).
From \fref{fig:vaa}(A), the CD potential for $l_x\ge4$ (even parity) is confirmed to be identical to the DF potential.

Coulomb potentials $V_{\rm C}(\bar{x}_i)$ are compared in \fref{fig:vaa}(B).
The styles of curves are the same as those in \fref{fig:vaa}(A).
The Coulomb potential for AB is calculated from the point + uniform charge distribution of the radius 2.94 fm \cite{Ngu13}.
For CD, it is calculated with two uniform charge spheres \cite{Dev75}.
The radii are set to 2.1 fm so as to make an equivalent potential to the folded one.
The Coulomb potentials are independent of the parity and $l_{x_i}$.

\begin{figure}[t] 
  \begin{center}
  \begin{tabular}{cc}
    \includegraphics[width=0.4\linewidth]{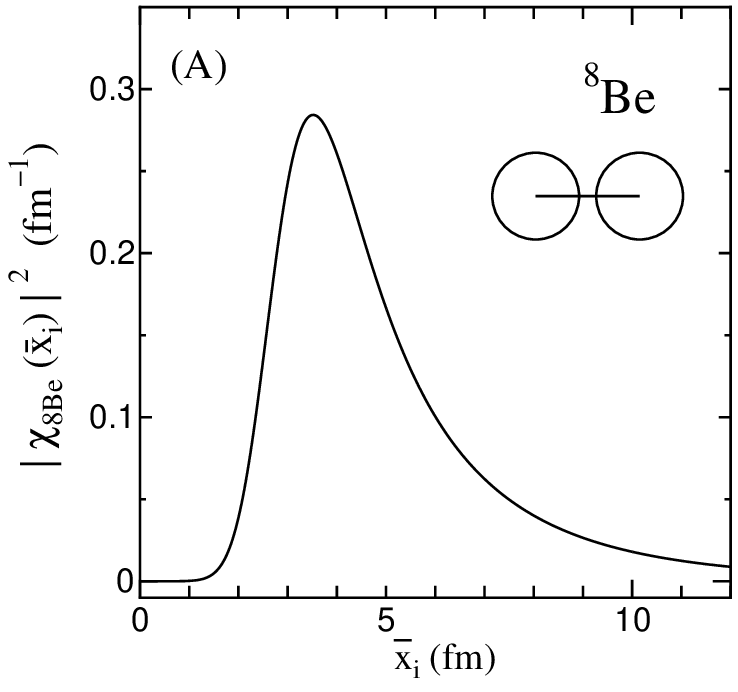} &
    \includegraphics[width=0.4\linewidth]{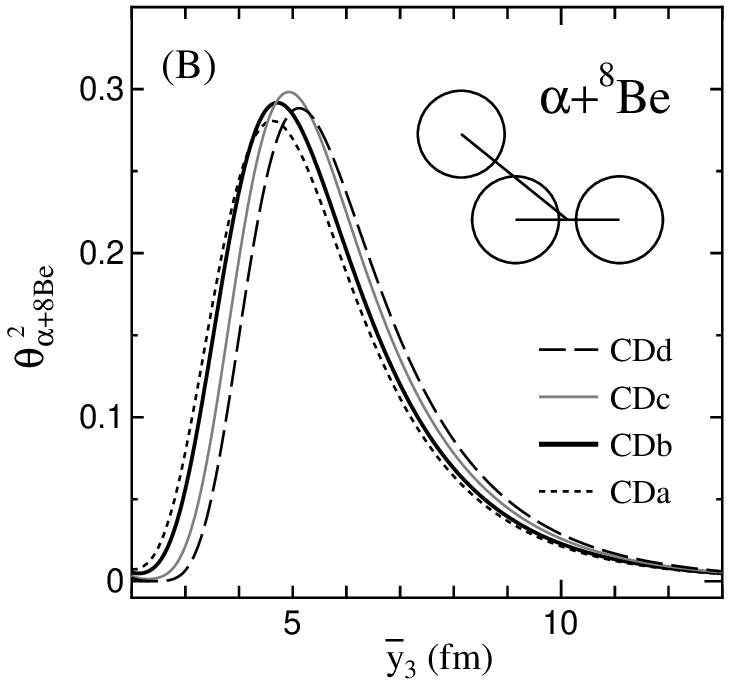}
  \end{tabular}
  \end{center}
  \caption{\label{fig:Be8}
(A) Density distribution of relative motion between $\alpha$+$\alpha$ in $^8$Be(0$^+_1$), obtained from the CD potential.
The most probable relative distance is found to be 3.52 fm.
The resonance energy $E_{\rm R}= 91.3$ keV and width $\Gamma=5.79$ eV are derived from the $s$-wave phase shifts.
(B) Dimensionless reduced width for $\alpha+^8$Be in $^{12}$C(0$^+_2$), defined in (\ref{eq:arw}).
The dotted, bold solid, thin solid and dashed curves are obtained from CDa, CDb, CDc and CDd, respectively.
The dilute structure is developed as $r_{3\alpha}$ increases.
  }
\end{figure}

The calculated resonant energy and width for $^8$Be(0$^+_1$) from CD are $E_{\rm R}=91.3$ keV and $\Gamma=5.79$ eV.
These are obtained from an $R$-matrix fit to the calculated $s$-wave phase shifts, and they are comparable with the experimental values \cite{Til04}: $E_{\rm R}=91.8$ keV and $\Gamma=5.57\pm0.25$ eV.
The corresponding wavefunction of relative motion between $\alpha$+$\alpha$ is depicted in \fref{fig:Be8}(A), $\chi_{^8{\rm Be}}=\bar{x}_i\Psi_{^8{\rm Be}}$.
From the solid curve, the most probable relative distance is expected to be 3.52 fm.
This wavefunction is used in (\ref{eq:gamma_alpha}) to estimate $\theta^2_{\alpha+^8{\rm Be}}$, which is displayed in \fref{fig:Be8}(B).

CD is similar to AB(D) used in CMF.
However, AB(D) makes $E_{\rm R}= 95.1$ keV and $\Gamma= 8.32$ eV for $^8$Be(0$^+_1$)~\cite{Ish13}.
They do not seem to be better than those of CD.
The nucleon density distribution in DF is calculated from the intrinsic wavefunction of $\alpha$-particle.
This means that the internal structure of $^4$He is included in the $\alpha$+$\alpha$ potential.
To take account of the microscopic aspect, I adopt CD as an equivalent local potential, updating the parameters to make a better reproduction of the experimental data.
In addition, AB used in ACF is assumed to be independent of $l_x$, i.e., ACF cannot even reproduce the $l_x=2$ phase shifts of $\alpha$+$\alpha$.

\subsection{3$\alpha$ potentials}\label{sec:4.2}

\begin{table}[t] 
    \caption{\label{tb:cdv3}
Variations in the quantities for the generated states to the range of 3$\alpha$ potentials.
$r_{3\alpha}$ and $v_{3\alpha,l}$ are defined in (\ref{eq:v3a}).
$E(l^\pi_i)$ is the energy of states with $l^\pi$.
$q_{\rm mix}$ is a mixing ratio of $\alpha+^8$Be clustering in 0$^+_2$.
The rms radius of 0$^+_2$, defined as (\ref{eq:rms}), is also listed.
${\cal N}^2$ is defined in (\ref{eq:2+}).
    }
    \begin{indented}
    \lineup
    \item[]\begin{tabular}{@{}*{7}{l}}
      \br
                      &       & \0\0CDa     &  \0CDb    &  \0CDc   &  \0CDd    & \0\0AB \\
      $r_{3\alpha}$     &(fm)   &\m\0\03.46   &\m\04.0    &\m\05.0   &\m\06.0    &\m\06.0 \\
      \mr
      $v_{3\alpha,0}$    & (MeV)&$-154.55$    &$-86.965$  &$-37.15$   &$-19.447$  &$-20.145$ \\
      $v_{3\alpha,2}$    & (MeV)&\0$-82.06$   &$-44.49$   &$-21.14$   &$-13.43$   &$-16.36$  \\
      $E(0^+_2)$       &(MeV) &\m\0\00.3794 &\m\00.3796 &\m\00.3796 &\m\00.3795 &\m\00.3795 \\
      $E(2^+_1)$       &(MeV) &\0\0$-2.837$ &\0$-2.836$ &\0$-2.836$ &\0$-2.836$ &\0$-2.838$ \\
      $E(0^+_1)$       &(MeV) &\0\0$-7.726$ &\0$-8.005$ &\0$-7.213$ &\0$-6.152$ &\0$-5.187$ \\
      $q_{\rm mix}$      &      &\m\0\00.52   &\m\00.56   &\m\00.63   &\m\00.65   &\m\00.61   \\
      $R_{\rm rms}(0^+_2)$&(fm) &\m\0\03.35   &\m\03.41    &\m\03.52   &\m\03.59   &\m\03.75  \\
      ${\cal N}^2$     &      &\m\0\00.784  &\m\00.928  &\m\01.054  &\m\01.863  &\m\01.637  \\
      \br
    \end{tabular}
    \end{indented}
\end{table}

The 3$\alpha$ potentials in (\ref{eq:ccpot}) are defined, as follows:
\begin{eqnarray}
  V^l_{3\alpha}(\rho) &=& v_{3\alpha, l} \exp\left[-\left(\frac{\rho}{r_{3\alpha}}\right)^2\right].
  \label{eq:v3a}
\end{eqnarray}
$v_{3\alpha, l}$ and $r_{3\alpha}$ are the parameters of the 3$\alpha$ potentials.
$v_{3\alpha, l}$ are adjusted with a given $r_{3\alpha}$ so as to minimize a difference from the experimental energies of 0$^+_2$ and 2$^+_1$ in $^{12}$C \cite{Ngu13,Ish13,Sun16,Fed96}.
The resultant values and calculated $E(l^\pi_i)$ are listed in \tref{tb:cdv3}.
Including the 3$\alpha$ potential, AB is basically the same as the potentials in \cite{Ngu13}.
For CD, I examine $r_{3\alpha}= 3.46$, 4.0, 5.0 and 6.0 fm.
In the present article, they are referred to as CDa, CDb, CDc and CDd, respectively, along with the $\alpha$+$\alpha$ potential.
CDa and CDb may resemble AB(D)+$\Delta(3.45)$ and AB(D)+$\Delta(3.9)$, recommended in \cite{Ish16}.
CDd and AB seem to generate the slightly high $E(0^+_1)$.

\subsection{Density distribution functions of 0$^+_2$ and 2$^+_1$ in $^{12}$C}\label{sec:4.3}

\begin{figure}[t] 
  \begin{center}
  \begin{tabular}{cc}
    \includegraphics[width=0.46\linewidth]{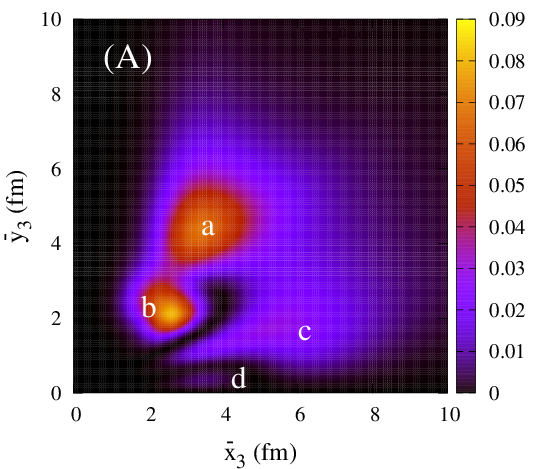} &
    \includegraphics[width=0.46\linewidth]{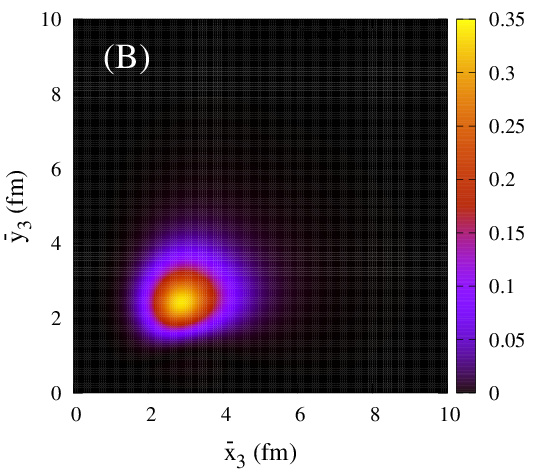} 
  \end{tabular}
  \end{center}
  \caption{\label{fig:prob-CD}
Density distribution functions of (A) 0$^+_2$ and (B) 2$^+_1$, defined in (\ref{eq:dens}), calculated from HHR$^\ast$ with CDb.
The horizontal and vertical axes are $\bar{x}_3$ and $\bar{y}_3$ in \fref{fig:jacobi}.
The results of HHR$^\ast$ seem to be similar to those of HHR \cite{Ngu13}, CMF \cite{Ish14} and \cite{Mor21}.
  }
\end{figure}

\Fref{fig:prob-CD}(A) shows the density distribution function of 0$^+_2$ obtained from CDb.
This color contour plot seems similar to that from HHR \cite{Ngu13}, CMF \cite{Ish14} and \cite{Mor21}.
Two conspicuous peaks are located on ($\bar{x}_3$, $\bar{y}_3$) = a(3.4 fm, 4.4 fm) and b(2.6 fm, 2.1 fm), and a plateau is found at long $\bar{x}_3$ and short $\bar{y}_3$.
A weak islet also appears at ($\bar{x}_3$, $\bar{y}_3$) = d(3.5 fm, 0.4 fm).
The peak (a) has an $\alpha+^8$Be configuration because the relative distance $\bar{x}_3$ in $^{12}$C(0$^+_2$) is close to that in $^8$Be(0$^+_1$).
The peak (b) corresponds to a configuration of an equilateral triangle with sizes approximately 2.6 fm.
The plateau (c) is an obtuse triangle, and the height at (c) is half of (a), ${\cal P}_{0^+_2}(c)={\cal P}_{0^+_2}(a)/2$ \cite{Ish14}.
(d) is almost linear alignment.
Because (c) is geometrically associated with (a), the 0$^+_2$ state is confirmed to have dual aspects: (a) dilute $\alpha$+$^8$Be and (b) spatially-shrunk structure corresponding to quantum liquid \cite{Ots22,Toh01}.
In addition, it is confirmed to decay dominantly via $\alpha+^8$Be because of the penetrability.

The similar density distributions of 0$^+_2$ can be obtained with CDa, CDc, CDd and AB.
However, the peak values are slightly varied.
To assess the differences, I define a mixing ratio of (a) dilute $\alpha+^8$Be to (b) spatial-shrunk structure, $q_{\rm mix}\equiv 1-{\cal P}_{0^+_2}(b)/[1.5\times{\cal P}_{0^+_2}(a)+{\cal P}_{0^+_2}(b)]$, as a guide.
From $q_{\rm mix}$ in \tref{tb:cdv3}, the dilute structure is found to develop as $r_{3\alpha}$ increases in CD.
Especially, CDc and CDd may be preferable from 61\% \cite{Ots22} clustering in 0$^+_2$.
This feature can be also seen in $\theta^2_{\alpha+^8{\rm Be}}$ of \fref{fig:Be8}(B).
As $r_{3\alpha}$ increases, the peak of $\theta^2_{\alpha+^8{\rm Be}}$ is shifted to large $\bar{y}_3$, and $\alpha$-particle probability increases for $\bar{y}_3>5$ fm.
$R_{\rm rms}(0^+_2)$ also becomes larger, as listed in \tref{tb:cdv3}.
The differences between CDd and AB may stem from the $\alpha$+$\alpha$ potential.

For 2$^+_1$, single peak is found in the density distribution function obtained from CDb (\fref{fig:prob-CD}(B)).
This corresponds to a spatial-shrunk configuration of an equilateral triangle with sizes approximately 2.8 fm.
The similar contour plots can be obtained with CDa, CDc, CDd and AB.
${\cal N}^2$ of (\ref{eq:2+}), obtained within a factor of 2 (\tref{tb:cdv3}), are consistent with those in HHR, CMF and ACF.
Especially, CDb and CDc seem to be preferable.
The ground state also has the same configuration as 2$^+_1$.

\begin{table}[t] 
    \caption{\label{tb:states}
Comparison in $E(l^\pi_i)$, $R_{\rm rms}(l^\pi_i)$, $B(E2)$, $M(E0)$ and $\Gamma(0^+_2)$.
The values of HHR$^\ast$ are the present results with CDb, and they appear to be concordant with the results of HHR \cite{Ngu13}, CMF \cite{Ish13}, ACF \cite{Sun16} and the experimental data \cite{Kel17}.
    }
    \begin{indented}
    \lineup
    \item[]\begin{tabular}{@{}*{7}{l}}
      \br
      & &  HHR$^\ast$ &  HHR \cite{Ngu13} & CMF \cite{Ish13} & ACF \cite{Sun16} & Exp. \cite{Kel17} \\
      \mr
      $E(0^+_2)$                      &(MeV)      &\m\00.3796 & \m0.380  &\m0.378   &\m\00.379    &\m\00.379\\
      $E(2^+_1)$                      &(MeV)      &\0$-2.836$ & $-2.875$ & $-2.83$  &\0$-2.836$  &\0$-2.835$\\
      $E(0^+_1)$                      &(MeV)      &\0$-8.005$ & \0\0---  & $-7.789$ &\0$-9.242$  &\0$-7.275$\\
      $R_{\rm rms}(0^+_2)$              &(fm)        &\m\03.41  & \0\0---  &\m3.43    &\m\04.00     &\m\0\0--- \\
      $R_{\rm rms}(2^+_1)$              &(fm)        &\m\02.39  &  \m2.459 &\0\0---   &\m\02.40     &\m\0\0--- \\
      $R_{\rm rms}(0^+_1)$              &(fm)        &\m\02.36  & \0\0--- &\0\0---    &\m\02.30     &\m\02.35--2.48\\
      $B(E2;0^+_2\!\rightarrow 2^+_1)$&($e^2$fm$^4$)&\m15.2    & \0\0--- & \m8.7     &\m34.6       &\m13.8 \\
      $B(E2;2^+_1\!\rightarrow 0^+_1)$&($e^2$fm$^4$)&\m\08.00  & \0\0--- &\0\0---    &\m12.4       &\m\07.76\\
      $M(E0;0^+_2\!\rightarrow 0^+_1)$&($e\,$fm$^2$)&\m\05.54  & \0\0--- &\0\0---    &\m\06.44     &\m\05.48\\
      $\Gamma(0^+_2)$                 &(eV)        &\m\07.1   & \0\0--- & \m6.9     &\m15.8       &\m\09.3\\
      \br
    \end{tabular}
    \end{indented}
\end{table}

Considering the variations to $r_{3\alpha}$ discussed above, I calculate the recommended values from CDb, and show the uncertainties of the results by using CDa and CDc in \sref{sec:5}.

In \tref{tb:states}, $E(l^\pi_i)$, $R_{\rm rms}(l^\pi_i)$, $B(E2)$, $M(E0)$ and $\Gamma(0^+_2)$ of CDb are compared with those of \cite{Ngu13,Ish13,Sun16,Kel17}.
The present results appear to be concordant with the previous values.
$R_{\rm rms}(0^+_2)$ is longer than $R_{\rm rms}(0^+_1)$, whereas $R_{\rm rms}(2^+_1)$ is similar to $R_{\rm rms}(0^+_1)$.
The derived monopole matrix element of (\ref{eq:ME0}) is 5.54 $e\,$fm$^2$.
The calculated charge radius of $^{12}$C(0$^+_1$) is 2.49 fm, comparable to the experimental value 2.48 fm \cite{Kel17}.
The $\alpha$+$^8$Be width, given by (\ref{eq:wa8Be}), is 2.37 eV at $\bar{y}_3= 6$ fm.
Thus, the calculated $\alpha$-decay width is obtained as 7.1 eV, and it appears to be consistent with \cite{Kel17}.

\section{Results: Triple-$\alpha$ reaction rates}\label{sec:5}

\begin{figure}[t] 
  \begin{center}
    \includegraphics[width=0.64\linewidth]{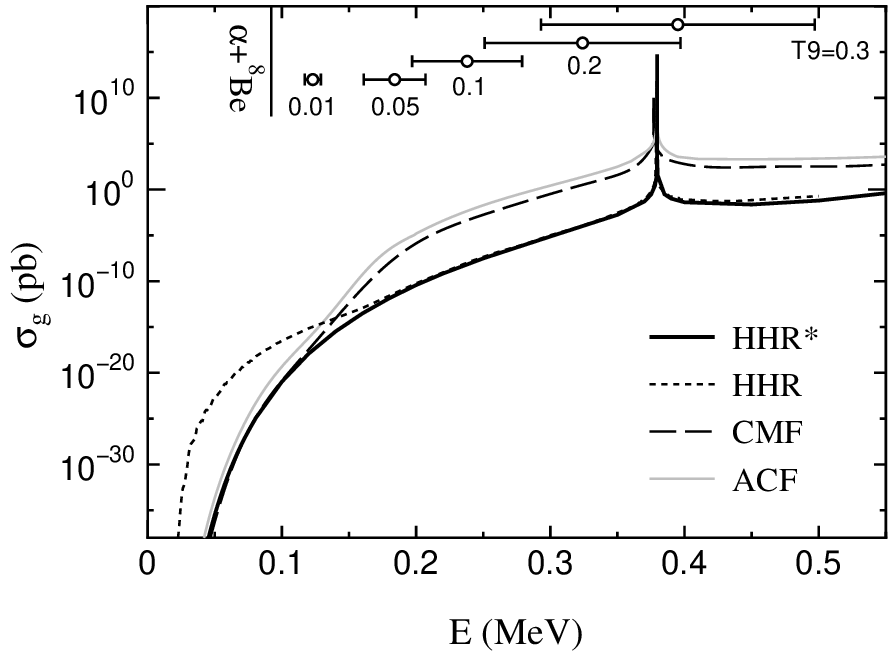}
  \end{center}
  \caption{\label{fig:sig}
Cross sections for photo-disintegration of $^{12}$C(2$^+_1 \rightarrow$ 0$^+$), defined by \eref{eq:sig}.
The solid curve is the result obtained from HHR$^\ast$ with CDb.
The dotted, dashed and gray curves are taken from HHR \cite{Ngu13}, CMF \cite{Ish13} and ACF \cite{Sun16}, respectively.
The Gamow peak energy \cite{Nacre} and energy-window \cite{Nacre} for $\alpha$+$^8$Be at $T_9= 0.01$, 0.05, 0.1, 0.2 and 0.3 are displayed by the open circles and horizontal bars as a guide of the energy range for the sequential process.
  }
\end{figure}

In this section, I discuss the calculated results of the triple-$\alpha$ reaction rates, after illustrating the photo-disintegration of $^{12}$C(2$^+_1 \rightarrow 0^+$) and the $S$-factor.
At the same time, I discuss the difference between adiabatic and non-adiabatic approaches.
The derived rates are expressed in an analytic form, and they are provided in a tabular form.
They are also converted into REACLIB format.
I discuss an update of the evaluated reaction rates, and I finally assess the present rates briefly.

\subsection{Comparison between photo-disintegration cross sections}\label{sec:5.1}

\begin{figure}[t] 
  \begin{center}
  \begin{tabular}{cc}
    \includegraphics[width=0.45\linewidth]{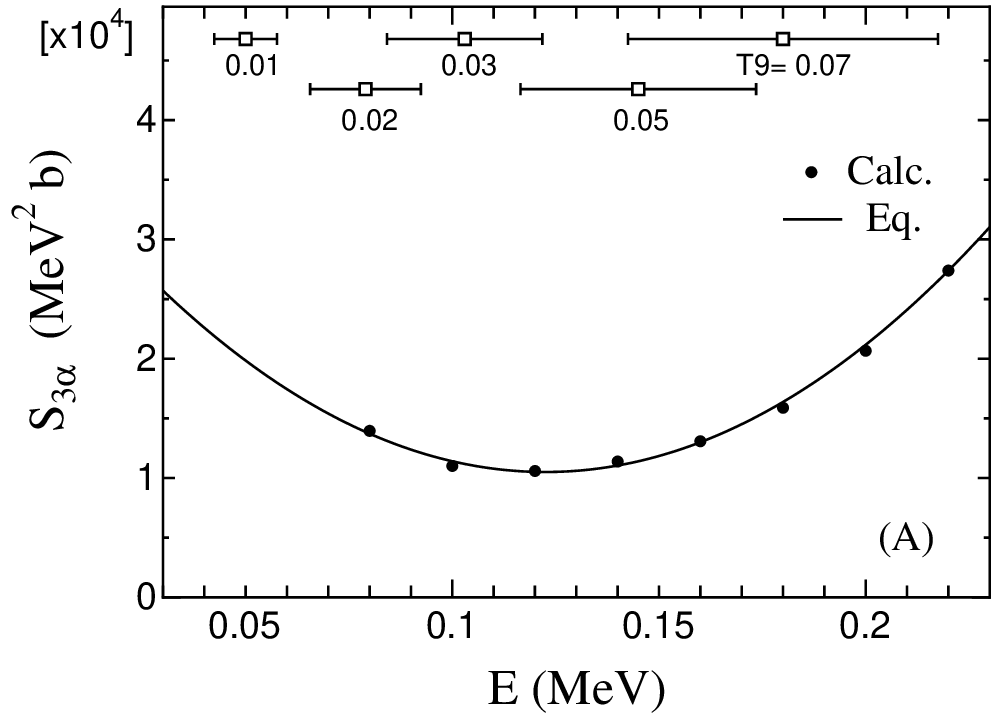} &
    \includegraphics[width=0.45\linewidth]{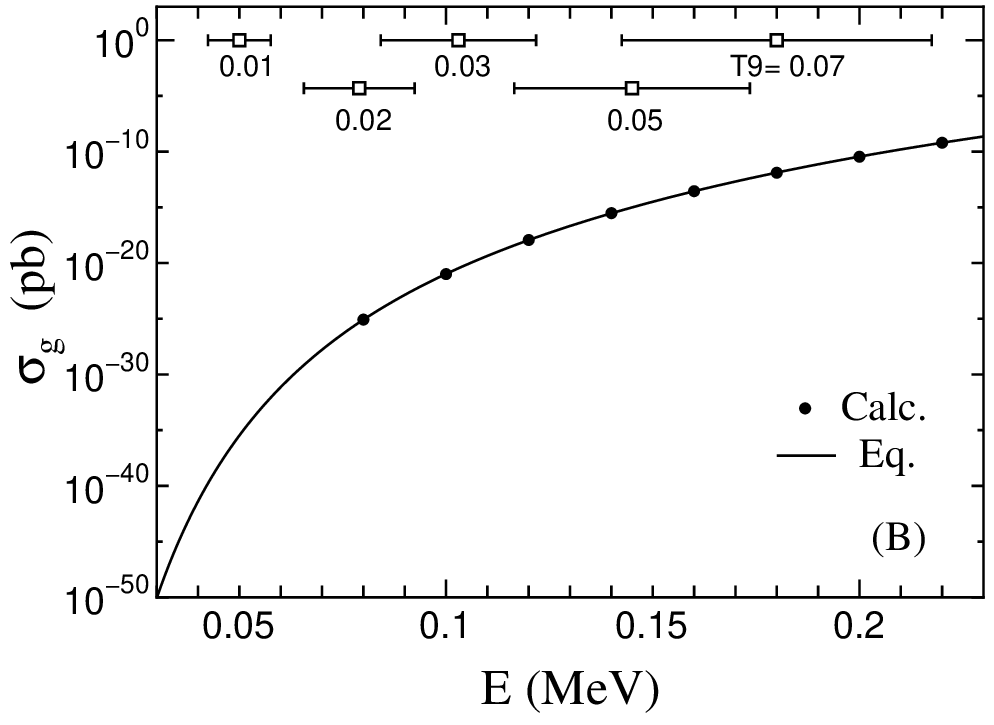}
  \end{tabular}
  \end{center}
  \caption{\label{fig:sfact}
(A) $S$-factors and (B) cross sections for photo-disintegration of $^{12}$C($2^+_1\rightarrow 0^+$).
The solid circles are the calculated values from HHR$^\ast$ with CDb.
The solid curves are (A) $S_{3\alpha}(E)= 3.717\times10^4 \cdot (1-11.71 E+ 47.79 E^2)$ MeV$^2$~b and (B) $\sigma_{\rm g}(E)= S_{3\alpha}(E)(E+2.836)^{-2}\exp(-26.125/\sqrt{E}-5E)$~b.
The open squares and horizontal bars are the Gamow peak energies and energy-windows at the respective temperatures for the direct triple-$\alpha$ process, defined in (\ref{eq:GE}) and (\ref{eq:GW}).
  }
\end{figure}

The photo-disintegration cross sections of $^{12}$C(2$^+_1 \rightarrow 0^+$), calculated from HHR$^\ast$ with CDb, are shown by the solid curve in \fref{fig:sig}.
A prominent narrow resonance of 0$^+_2$ is found at $E\approx 0.379$ MeV, and the smoothly varying non-resonant cross sections are obtained at off-resonant energies.
For $0.15 \le E \le 0.35$ MeV, I find $\sigma_{\rm g} = 10^{-15}$--$10^{-3}$ pb order of magnitude.
This result appears to be almost identical to HHR \cite{Ngu13} shown by the dotted curve.
Below $E=0.15$ MeV, $\sigma_{\rm g}$ of HHR$^\ast$ seems similar to that of CMF (dashed curve) and ACF (gray curve).
On the other hand, the present results at off-resonant energies above $E=0.15$ MeV are 4 orders of magnitude smaller than the values predicted by CMF and ACF.
As described in \sref{sec:1}, CMF has been developed in the three-body bound states and p+d reactions.
The internal motion of $^8$Be in break-up channels is described as an approximation within a certain range \cite{Ish13}.
For ACF, the adiabatic potential, generated separately for a fixed hyper-radius, is characterized by $\alpha+^8$Be at short radii and by 3$\alpha$ at large radii, and their $\sigma_{\rm g}$ for $E\ge 0.13$ MeV has been interpreted as a part of the 0$^+_2$ resonance \cite{Sun16}.
Judging from their theoretical approaches, most of the differences above $E=0.15$ MeV seem to stem from the internal adiabatic feature.
The adiabatic approach to $^8$Be continuum states makes the long resonant tail of 0$^+_2$, leading to the sequential decay process at off-resonant energies,
and it might have enhanced the photo-disintegration cross sections unexpectedly.
The enhancement of \cite{Ngu13} for $E<0.15$ MeV seems to be caused by the redundant global backward propagation of (\ref{eq:bwprop}) and the difference in the Coulomb potentials of (\ref{eq:zeff}).

The Gamow peak energy and energy-window for $\alpha+^8$Be are given by $E_{\alpha{\rm ^8Be}} = 0.677 T_9^{2/3}+0.092$ MeV \cite{Nacre} and $\Delta_{\alpha{\rm ^8Be}} = 0.558 T_9^{5/6}$ \cite{Nacre}, and they are shown by the open circles and horizontal bars in \fref{fig:sig}.
From the difference above $E=0.15$ MeV, the reaction rate of HHR$^\ast$ at $T_9=0.05$ is expected to be 4 orders of magnitude smaller than that of CMF and ACF.

To examine the long-range Coulomb couplings, I also execute the calculations without the screening potential.
The differences, defined as (\ref{eq:diff_scrn}), are $d_{\rm scrn}=$ 0.75\%, 0.34\%, 0.31\%, 0.30\% and 0.07\% at $E=0.2$, 0.3, 0.3796, 0.4 and 0.5 MeV.
I therefore find that the differences are less than 1\% at the energies corresponding to helium burning temperatures, and that the Coulomb couplings seem to be negligible for $\rho>650$ fm in practice.

\begin{table}[t] 
    \caption{\label{tb:sfact}
Coefficients in the $S$-factors of (\ref{eq:3a_fact}).
The resonant energy and $\gamma$ width of 0$^+_2$ are also listed.
The triple-$\alpha$ reaction rates for 0$^+$ are obtained from these quantities with (\ref{eq:rates_nr}), (\ref{eq:rates_st}) and (\ref{eq:rates_r}).
The recommended rates are obtained from CDb.
    }
    \begin{indented}
    \item[]\begin{tabular}{@{}*{8}{c}}
      \br
       &$s_0$ & $s_1$ & $s_2$ & $\eta_0$ & $a$ & $E(0^+_2)$ & $\Gamma_\gamma(0^+_2)$ \\
       &(MeV$^2$~b) & (MeV$^{-1}$) & (MeV$^{-2}$) & (MeV$^{1/2}$) & (MeV$^{-1}$) & (MeV) & (meV) \\
      \mr
      CDa &1.416$\times10^4$ & $-11.18$ & 43.87 & 4.068 & 5.0 & 0.3794 & 3.9 \\
      CDb &3.717$\times10^4$ & $-11.71$ & 47.79 & 4.158 & 5.0 & 0.3796 & 3.9 \\
      CDc &3.075$\times10^4$ & $-11.71$ & 48.83 & 4.152 & 5.0 & 0.3796 & 3.9 \\
      CDd &3.525$\times10^4$ & $-11.60$ & 48.40 & 4.129 & 5.0 & 0.3795 & 3.9 \\
      AB  &9.166$\times10^4$ & $-11.66$ & 48.08 & 4.125 & 5.0 & 0.3795 & 3.9 \\
      \br
    \end{tabular}
    \end{indented}
\end{table}

The calculated $\sigma_{\rm g}$ below $E=0.2$ MeV can be expressed in terms of $S_{3\alpha}$.
The solid circles in \fref{fig:sfact}(A) are the calculated values of HHR$^\ast$ with CDb, and the solid curve is $S_{3\alpha}(E) = 3.717\times10^4 \cdot (1-11.71 E + 47.79 E^2 )$ MeV$^2$~b.
Compared with $\sigma_{\rm g}$ in \fref{fig:sfact}(B), the energy dependence of $S_{3\alpha}$ is quite weak.
The open squares and horizontal bars are the Gamow peak energies and energy-windows for 3$\alpha$, defined in (\ref{eq:GE}) and (\ref{eq:GW}).
In the present article, the reaction rates below $T_9 \approx 0.02$ are generated from the extrapolated $S_{3\alpha}$, assuming that $S_{3\alpha}$ has weak variation to $E$.
The cross sections from CDa, CDc, CDd and AB are also expressed with $S_{3\alpha}$ similarly well.
The resultant coefficients are summarized in \tref{tb:sfact}.
From these quantities, the triple-$\alpha$ reaction rates for 0$^+$ are given with (\ref{eq:rates_nr}), (\ref{eq:rates_st}) and (\ref{eq:rates_r}).

\subsection{Triple-$\alpha$ reaction rates}\label{sec:5.2}

\begin{figure}[t] 
  \begin{center}
  \begin{tabular}{cc}
    \includegraphics[width=0.42\linewidth]{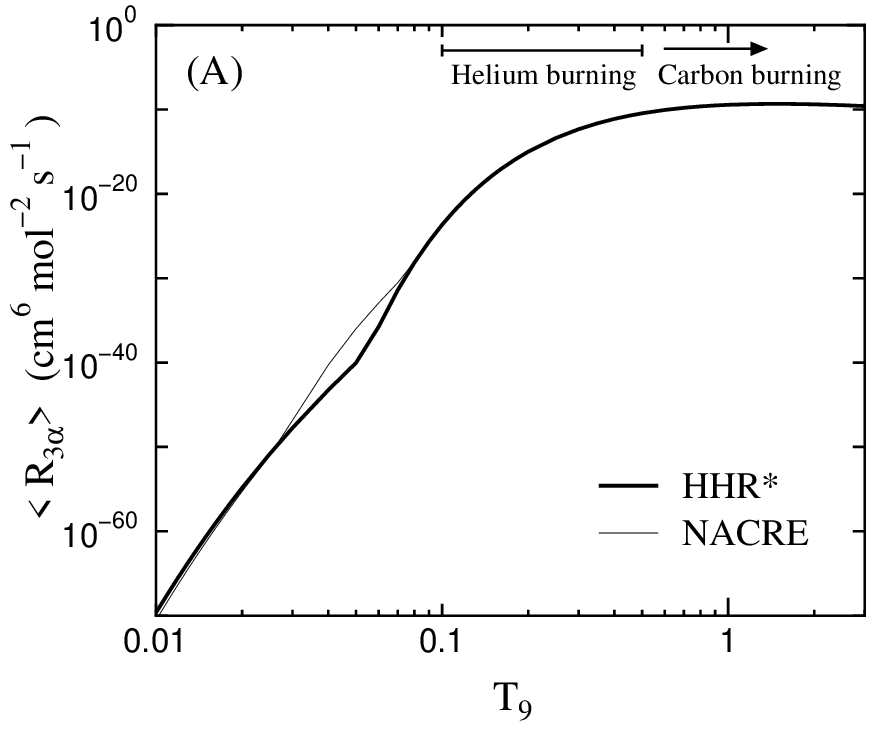} &
    \includegraphics[width=0.42\linewidth]{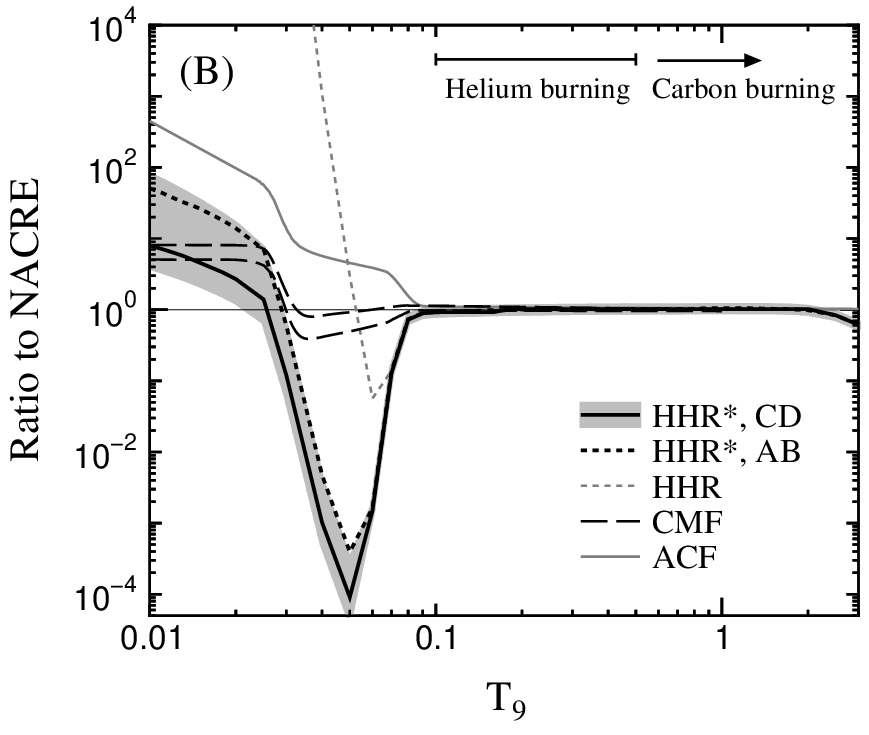}
  \end{tabular}
  \end{center}
  \caption{\label{fig:rates}
Triple-$\alpha$ reaction rates (A) in units of cm$^6$mol$^{-2}$s$^{-1}$ and (B) in ratio to NACRE.
The bold solid and bold dotted curves are the result obtained from HHR$^\ast$ with CDb and AB.
The dotted, dashed and gray curves are taken from HHR \cite{Ngu13}, CMF \cite{Ish13} and ACF \cite{Sun16}, respectively.
The thin solid curve is the NACRE reaction rates \cite{Nacre}.
The horizontal bars are the range of helium burning temperatures.
The gray shade represents the uncertainties of the present rates.
  }
\end{figure}

The resulting reaction rates are shown in \fref{fig:rates}: (A) in units of cm$^6$mol$^{-2}$s$^{-1}$ and (B) in ratio to NACRE.
The bold solid curves are the result obtained from CDb, and they are expressed as
\begin{eqnarray}
  \langle R_{3\alpha}\rangle
  &=&
  \Big[
   \frac{229.7}{T_9^{13/6}(1+ b T_9)^{5/6}}
  +\frac{2.5}{T_9^{11/6}(1+ b T_9)^{7/6}}
  -\frac{2911.}{T_9^{3/2}(1+ b T_9)^{3/2}}
  \nonumber\\
  &&
  -\frac{225.}{T_9^{7/6}(1+ b T_9)^{11/6}}
  +\frac{1.285\times10^4}{T_9^{5/6}(1+ b T_9)^{13/6}}
  +\frac{2.53\times10^3}{T_9^{1/2}(1+ b T_9)^{5/2}}
  \Big]
  \nonumber\\
  &\cdot& \exp\Big[-\frac{37.673}{T_9^{1/3}}(1+b T_9)^{1/3}\Big]
  + \frac{2.966\times10^{-8}}{T_9^3}\exp\Big(-\frac{4.4053}{T_9}\Big),
  \label{eq:CDb}
  \nonumber\\
  &&
\end{eqnarray}
where $b=0.431$.
As shown in this figure, the derived rates are found to be consistent with NACRE at helium burning temperatures (horizontal bars).
Especially, the rates are approximately proportional to $T_9^{41}$ at $T_9=0.1$, as are the results in NACRE, so that they are expected to make a red-giant consistently in the late stage of stellar evolution.

On the other hand, the present result is also found to be reduced by a factor of 10$^{-4}$ at $T_9 = 0.05$.
The dashed and gray curves are taken from CMF \cite{Ish13} and ACF \cite{Sun16}.
As shown in \fref{fig:sig}, the present rate is much smaller than that of CMF and ACF at $T_9=0.05$ because $\sigma_{\rm g}$ is reduced from theirs at $E = 0.18$ MeV.
Due to the strong influence of 0$^+_2$, the large difference in $\sigma_{\rm g}$ above $E=0.20$ MeV does not make a large deviation in the rates above $T_9=0.08$.
The similar reduction can be found in the rates from AB, shown by the bold dotted curve.
From a comparison between the bold solid and bold dotted curves, I confirm that the reduction at $T_9=0.05$ does not stem from the adopted $\alpha$+$\alpha$ potentials.

The uncertainties of the rates are shown by the gray shade in \fref{fig:rates}(B).
The upper and lower limits are defined as $\langle R_{3\alpha} \rangle_{\rm U}= 1.1\times\langle R_{3\alpha}\rangle_a$ and $\langle R_{3\alpha} \rangle_{\rm L}= 0.9\times\langle R_{3\alpha}\rangle_c$ for $T_9\ge0.06$, including the experimental uncertainty of $\Gamma_\gamma(0^+_2)$.
For $T_9\le0.05$, they are assumed as $\langle R_{3\alpha} \rangle_{\rm U}= 2 \times\langle R_{3\alpha}\rangle_a$ and $\langle R_{3\alpha} \rangle_{\rm L}= \langle R_{3\alpha}\rangle_c/2$.
$\langle R_{3\alpha}\rangle_a$ and $\langle R_{3\alpha}\rangle_c$ are the rates obtained from CDa and CDc, respectively.
The numerical values of the resultant rates are listed in \tref{tb:rates}.

\begin{landscape}
\begin{table} 
    \caption{\label{tb:rates}
Triple-$\alpha$ reaction rates by HHR$^\ast$.
The rates are given in units of cm$^6$mol$^{-2}$s$^{-1}$.
The recommended rates $\langle R_{3\alpha}\rangle$ are obtained from CDb, and they are analytically expressed with (\ref{eq:CDb}).
$\langle R_{3\alpha} \rangle_{\rm L}$ and $\langle R_{3\alpha} \rangle_{\rm U}$ are the lower and upper limits, respectively.
The values of $\langle R_{3\alpha}\rangle_{\rm L}$, $\langle R_{3\alpha}\rangle$ and $\langle R_{3\alpha}\rangle_{\rm U}$ are multiplied by 10$^x$.
The ratios to CF88 \cite{CF88} and NACRE \cite{Nacre} are also listed for reference.
    }
    \begin{indented}
    \lineup
    \item[]\begin{tabular}{@{}*{7}{l}|*{7}{l}}
      \br
      \0$T_9$ & $\langle R_{3\alpha}\rangle_{\rm L}$ & $\langle R_{3\alpha}\rangle$ & $\langle R_{3\alpha}\rangle_{\rm U}$ & $x$ & CF88 &NACRE&
      \0$T_9$ & $\langle R_{3\alpha}\rangle_{\rm L}$ & $\langle R_{3\alpha}\rangle$ & $\langle R_{3\alpha}\rangle_{\rm U}$ & $x$ & CF88 &NACRE\\
      \mr
0.010 & 1.15 &\02.33 & 23.2  & $-$70 & 87.   & 7.9 & 0.15 & 1.40 &\01.55 &\01.73 & $-$18 & 1.0 & 0.97\\
0.011 & 2.08 &\04.23 & 39.1  & $-$68 & 79.   & 7.1 & 0.16 & 7.22 &\07.99 &\08.92 & $-$18 & 1.0 & 0.97\\
0.012 & 2.05 &\04.20 & 36.2  & $-$66 & 72.   & 6.4 & 0.18 & 1.08 &\01.20 &\01.33 & $-$16 & 1.0 & 0.98\\
0.013 & 1.24 &\02.54 & 20.6  & $-$64 & 66.   & 5.7 & 0.20 & 9.09 & 10.1  & 11.2  & $-$16 & 1.0 & 0.99\\
0.014 & 4.98 & 10.2  & 78.7  & $-$63 & 60.   & 5.1 & 0.25 & 3.81 &\04.22 &\04.69 & $-$14 & 1.1 & 1.0 \\
0.015 & 1.42 &\02.92 & 21.4  & $-$61 & 54.   & 4.6 & 0.30 & 4.16 &\04.61 &\05.11 & $-$13 & 1.1 & 1.0 \\
0.016 & 3.03 &\06.24 & 43.7  & $-$60 & 48.   & 4.1 & 0.35 & 2.13 &\02.36 &\02.62 & $-$12 & 1.1 & 1.0 \\
0.018 & 6.73 & 13.9  & 89.9  & $-$58 & 35.   & 3.3 & 0.40 & 6.89 &\07.64 &\08.45 & $-$12 & 1.1 & 1.0 \\
0.020 & 7.00 & 14.5  & 87.3  & $-$56 & 24.   & 2.7 & 0.45 & 1.64 &\01.82 &\02.02 & $-$11 & 1.1 & 1.0 \\
0.025 & 7.50 & 15.5  & 81.0  & $-$52 &\05.7  & 1.4 & 0.50 & 3.19 &\03.54 &\03.91 & $-$11 & 1.1 & 1.0 \\
0.03  & 8.80 & 18.2  & 84.2  & $-$49 &\00.29 & 0.12& 0.60 & 8.01 &\08.89 &\09.82 & $-$11 & 1.1 & 1.0 \\
0.04  & 2.71 &\05.55 & 21.1  & $-$44 &\01.9$\times10^{-3}$ & 1.0$\times10^{-3}$ & 0.70 & 1.44 &\01.60 &\01.76 & $-$10 & 1.1 & 1.0 \\
0.05  & 4.55 &\09.43 & 30.1  & $-$41 &\01.6$\times10^{-4}$ & 9.1$\times10^{-5}$ & 0.80 & 2.12 &\02.35 &\02.59 & $-$10 & 1.1 & 1.0 \\
0.06  & 1.65 &\01.81 &\02.08 & $-$36 &\02.6$\times10^{-3}$ & 1.5$\times10^{-3}$ & 0.90 & 2.74 &\03.05 &\03.36 & $-$10 & 1.1 & 1.0 \\
0.07  & 3.66 &\04.03 &\04.59 & $-$32 &\00.23 & 0.13 & 1.0 & 3.26 &\03.62 &\03.99 & $-$10 & 1.1 & 1.0 \\
0.08  & 6.39 &\07.05 &\07.98 & $-$29 &\00.92 & 0.73 & 1.25& 4.03 &\04.48 &\04.93 & $-$10 & 1.1 & 1.0 \\
0.09  & 2.04 &\02.25 &\02.54 & $-$26 &\01.0  & 0.89 & 1.5 & 4.20 &\04.66 &\05.13 & $-$10 & 1.1 & 1.0 \\
0.10  & 1.98 &\02.19 &\02.47 & $-$24 &\01.0  & 0.92 & 1.75& 4.02 &\04.47 &\04.92 & $-$10 & 1.1 & 1.0 \\
0.11  & 8.17 &\09.02 & 10.1  & $-$23 &\01.0  & 0.94 & 2.0 & 3.69 &\04.10 &\04.51 & $-$10 & 1.1 & 0.99\\
0.12  & 1.77 &\01.96 &\02.19 & $-$21 &\01.0  & 0.94 & 2.5 & 2.94 &\03.26 &\03.59 & $-$10 & 1.1 & 0.83\\
0.13  & 2.34 &\02.59 &\02.90 & $-$20 &\01.0  & 0.95 & 3.0 & 2.28 &\02.54 &\02.79 & $-$10 & 1.1 & 0.61\\
0.14  & 2.11 &\02.33 &\02.61 & $-$19 &\01.0  & 0.96 &     &      &       &       &       &     &     \\
      \br
    \end{tabular}
    \end{indented}
\end{table}
\end{landscape}

\subsection{Translation into REACLIB}\label{sec:5.3}

\begin{table}[t] 
    \caption{\label{tb:reaclib}
Coefficients of expansion in REACLIB format, obtained from HHR$^\ast$ with CDb.
The reaction rates are defined as (\ref{eq:reaclib}).
The maximum deviation from (\ref{eq:CDb}) is also listed.
    }
    \begin{indented}
    \lineup
    \item[]\begin{tabular}{@{}*{9}{l}}
      \br
      $i$ & \m\0$a_{i0}$ & \m$a_{i1}$ & \m\0$a_{i2}$& $a_{i3}$ & \m$a_{i4}$ & $a_{i5}$ &\m$a_{i6}$ & (\%) \\
      \mr
      1 & \m\02.4379 & \m0.         & $-37.219$ & 0.   & $-1.4$   & 0. & $-13/6$ & 14 \\
      2 & $-17.333$  & $-4.4053$    &  \m\00.   & 0.   & \m0.     & 0. & $-3.$  & ---  \\
      \br
    \end{tabular}
    \end{indented}
\end{table}

\begin{figure}[t] 
  \begin{center}
    \includegraphics[width=0.65\linewidth]{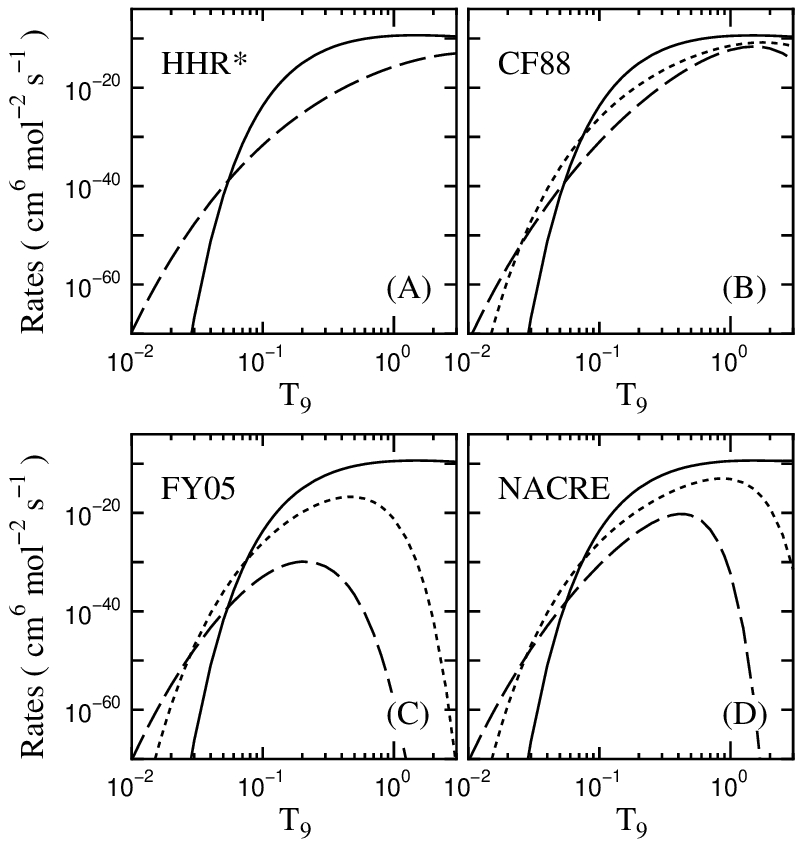}
  \end{center}
  \caption{\label{fig:reaclib}
Comparison with the triple-$\alpha$ reaction rates: (A) HHR$^\ast$, this work, (B) CF88, (C) FY05 and (D) NACRE.
HHR$^\ast$ consists of two components, whereas CF88, FY05 and NACRE consist of three components.
The solid and dashed curves represent the resonant and non-resonant components, respectively.
The dotted curves seem to originate from the assumed bound $^8$Be in CF88, FY05 and NACRE, and they are interpreted as the non-resonant sequential process between $\alpha$+$^8$Be.
HHR$^\ast$ does not have such a component.
The statistically generated $^8$Be is broken-up immediately by the third $\alpha$-particle at $T_9\approx0.05$.
  }
\end{figure}

The recommended rates of HHR$^\ast$, shown in the previous subsection, are converted into REACLIB format \cite{Reaclib}:
  \begin{eqnarray}
    \langle R_{3\alpha}\rangle &=&
    \mathop{\sum}_i \exp\big(a_{i0} +a_{i1}/T_9 +a_{i2}/T_9^{1/3}+a_{i3} T_9^{1/3}
    +a_{i4} T_9 \nonumber \\
    &+&a_{i5} T_9^{5/3} +a_{i6} \ln(T_9) \big).
    \label{eq:reaclib}
  \end{eqnarray}
$a_{ij}$ are the coefficients of expansion.
\Eref{eq:ex-rates_2} is adopted for the non-resonant component, and $a_{i4}$ is used for high temperatures.
The resultant $a_{ij}$ are listed in \tref{tb:reaclib}.
The maximum deviation from (\ref{eq:CDb}) is also listed.
The converted rates are displayed in \fref{fig:reaclib}(A).
The dashed ($i=1$) and solid ($i=2$) curves are the non-resonant and resonant components, respectively.
In REACLIB \cite{Reaclib}, two data sets, labeled with CF88 \cite{CF88} and FY05 \cite{Fyn05}, are compiled as the triple-$\alpha$ reaction rates.
They are also shown in figures~\ref{fig:reaclib}(B) and \ref{fig:reaclib}(C).
Both of them consist of three components.
The dashed and dotted curves contribute below $T_9=0.07$ as the non-resonant components, and the solid curves represent the resonant component.
Compared with CF88 and FY05, HHR$^\ast$ does not have a component of the dotted curve.

\begin{figure}[t] 
  \begin{center}
    \includegraphics[width=0.65\linewidth]{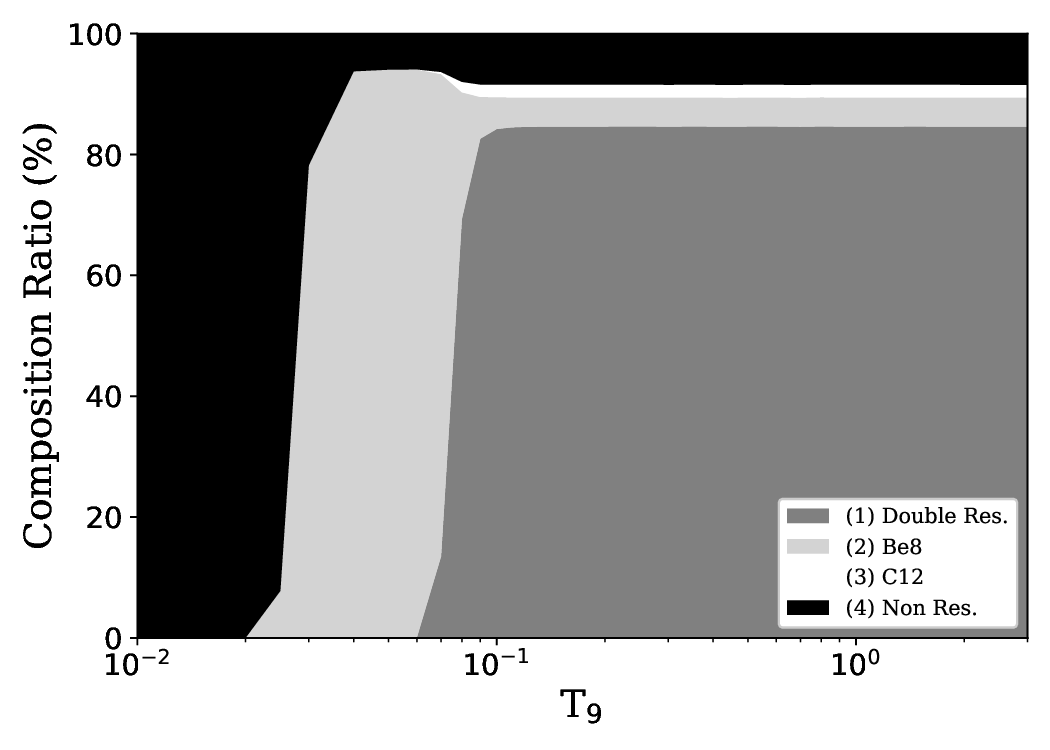}
  \end{center}
  \caption{\label{fig:composition}
Composition ratio of the triple-$\alpha$ reaction rates for 0$^+$ in NACRE \cite{Nacre}.
The rates are calculated from (\ref{eq:nacre-3a}) with the adopted experimental resonant energies and widths in \cite{Nacre}.
The dense and black areas are $\langle R_{3\alpha} \rangle_{\rm DR}/\langle R_{3\alpha} \rangle$ and $\langle R_{3\alpha} \rangle_{\rm NRs}/\langle R_{3\alpha} \rangle$, respectively.
The gray area represents $\langle R_{3\alpha} \rangle_{^8{\rm Be}}/\langle R_{3\alpha} \rangle$, the non-resonant sequential process between $\alpha$+$^8$Be.
The white area is $\langle R_{3\alpha} \rangle_{^{12}{\rm C}}/\langle R_{3\alpha} \rangle$, and it is small in the entire region.
The standard reaction rates are found to generally consist of three components.
  }
\end{figure}

To understand why the REACLIB rates have three components, let me recall the method used in NACRE.
In \cite{Nacre}, the triple-$\alpha$ reaction rates are given as
\begin{eqnarray}
  \langle R_{3\alpha} \rangle
  &=&
  \frac{24\hbar\,N_{\rm A}^2}{\pi(\mu_{12} \mu_{(12)3})^{1/2} (k_{\rm B} T)^3}\nonumber \\
  &\cdot&
  \int_0^\infty \!\!dE_1 \int_0^\infty \!\!dE_2
  \frac{\sigma_{\alpha\alpha}(E_1)\sigma_{\alpha ^8{\rm Be}}(E_2; E_1)}{\Gamma_\alpha(^8{\rm Be},E_1)}
  E_1 E_2 \exp\left( -\frac{E_1+E_2}{k_{\rm B} T}\right) \nonumber \\
  && \nonumber\\
  &=&
  \int_{E_{\rm R1}-\Delta\epsilon_1}^{E_{\rm R1}+\Delta\epsilon_1} \!\!\int_{E_{\rm R2}-\Delta\epsilon_2}^{E_{\rm R2}+\Delta\epsilon_2} \!\!\!\bigcirc\, dE_1dE_2
  +\int_{E_{\rm R1}-\Delta\epsilon_1}^{E_{\rm R1}+\Delta\epsilon_1} \!\!\int_{\rm other}                           \!\!\!\bigcirc\, dE_1dE_2
  \nonumber \\
  &+&
  \int_{\rm other}                             \!\int_{E_{\rm R2}-\Delta\epsilon_2}^{E_{\rm R2}+\Delta\epsilon_2} \!\!\!\bigcirc\, dE_1dE_2
  +\int_{\rm other}                            \!\int_{\rm other}                           \!\!\!\bigcirc\, dE_1dE_2 \nonumber \\
  && \nonumber\\
  &\equiv&
  (\langle R_{3\alpha} \rangle_{\rm DR} + \langle R_{3\alpha} \rangle_{^8{\rm Be}} + \langle R_{3\alpha} \rangle_{^{12}{\rm C}} + \langle R_{3\alpha} \rangle_{\rm NRs}) / 100.
  \label{eq:nacre-3a}
\end{eqnarray}
$\Gamma_\alpha(^8{\rm Be}, E_1)$ and $\sigma_{\alpha\alpha}$ are the energy-dependent $\alpha$-width of $^8$Be and elastic cross sections between $\alpha$+$\alpha$, respectively, and they give $^8$Be formation probability.
$\sigma_{\alpha ^8{\rm Be}}$ is the radiative capture cross sections of $\alpha+^8$Be.
$E_1$ is the center-of-mass energy of relative motion between $\alpha$+$\alpha$, and $E_2$ is the energy between $\alpha+^8$Be.
In the first line of (\ref{eq:nacre-3a}), the integral of $E_1$ can be divided into two: the $^8$Be resonance energy region $E_{\rm R1}\pm\Delta\epsilon_1=92.08\pm0.03$ keV and the other.
Likewise, the integral of $E_2$ can be divided into $E_{\rm R2}\pm\Delta\epsilon_2= 287.7\pm0.05$ keV and the other.
The resonant energy of $^{12}$C(0$^+_2$) $E_{\rm R2}$ is defined by the center-of-mass energy of $\alpha$+$^8$Be(0$^+_1$) \cite{Nacre}.
Consequently, the NACRE rates are expressed with four components in the second line, and they are expressed as (1) double resonances of $^8$Be and $^{12}$C, $\langle R_{3\alpha} \rangle_{\rm DR}$, (2) $^8$Be resonance, $\langle R_{3\alpha} \rangle_{^8{\rm Be}}$, (3) $^{12}$C resonance, $\langle R_{3\alpha} \rangle_{^{12}{\rm C}}$ and (4) non-resonant contributions, $\langle R_{3\alpha} \rangle_{\rm NRs}$, in the third line.
These composition ratios are shown in \fref{fig:composition}.
$\langle R_{3\alpha}\rangle_{\rm DR}/\langle R_{3\alpha}\rangle$ and $\langle R_{3\alpha}\rangle_{\rm NRs}/\langle R_{3\alpha}\rangle$ are shown by the dense and black areas, respectively.
$\langle R_{3\alpha}\rangle_{\rm DR}$ accounts for over 80\% above $T_9= 0.09$, and $\langle R_{3\alpha}\rangle_{\rm NRs}$ dominates the rates for $T_9\le 0.025$.
$\langle R_{3\alpha}\rangle_{^8{\rm Be}}/\langle R_{3\alpha}\rangle$ is shown by the gray area, and it is found to dominate the rates for $0.03 \le T_9 \le 0.07$.
$\langle R_{3\alpha}\rangle_{^{12}{\rm C}}/\langle R_{3\alpha}\rangle$, shown by the white area, is small over the entire region.
The rates from 2$^+_2$ \cite{Zim13} and 3$^-_1$ \cite{Tsu21} of $^{12}$C are small below $T_9=3$.
Therefore, the standard reaction rates below $T_9=3$ are found to generally consist of three components.
The analytic expansion of NACRE is shown in \fref{fig:reaclib}(D). (\ref{app:2})

If they are classified into the reaction process, $\langle R_{3\alpha} \rangle_{\rm DR}$ means the sequential process via two narrow resonances, and $\langle R_{3\alpha} \rangle_{\rm NRs}$ reminds us of the direct triple-$\alpha$ process.
The component of $\langle R_{3\alpha} \rangle_{^8{\rm Be}}$ is interpreted as the non-resonant sequential process between $\alpha$+$^8$Be.
However, this term may not seem to be realistic, because $^8$Be is unbound.
In other words, the dotted curves in \fref{fig:reaclib} seem to originate from the assumed $^8$Be in CF88, FY05 and NACRE.
From the present non-adiabatic calculation, the statistically generated $^8$Be is found to be broken-up immediately by the third $\alpha$-particle before its lifetime at $T_9\approx 0.05$.
My calculated results, therefore, indicate that the non-resonant sequential process could be at least deleted to update the reaction rates in NACRE and REACLIB.

\subsection{Brief estimate of the astrophysical impact}\label{sec:5.4}

As a brief assessment of the rates, I finally show the ignition density \cite{Nom85,Lan86} of helium burning in accreting white dwarfs of close binary systems.
The ignition density $\rho_{\rm ign}$ is defined by nuclear heating time scale $\tau_{\rm He} = C_{\rm p} T/{\epsilon_{3\alpha}}=10^6$ yr~\cite{Nom85} and energy generation rates of the triple-$\alpha$ reaction \cite{Ngu12,Ngu13},
\begin{eqnarray}
  \epsilon_{3\alpha} &=& N_{\rm A} \,\frac{Q}{6} \left(\frac{X_{\rm He}}{4}\right)^3 \,\rho_{\rm ign}^2\, f_{\rm sc} \langle R_{3\alpha}\rangle.
\end{eqnarray}
$Q$ is Q-value of the 3$\alpha$ reaction; $X_{\rm He}$ is helium mass fraction; $f_{\rm sc}$ is a factor of electron screening.
I adopt two types of electron screening labeled with I90 \cite{Ito90} and SV \cite{Sal69}.
$C_{\rm p}$ is specific heat at constant pressure.

\begin{figure}[t] 
  \begin{center}
    \includegraphics[width=0.6\linewidth]{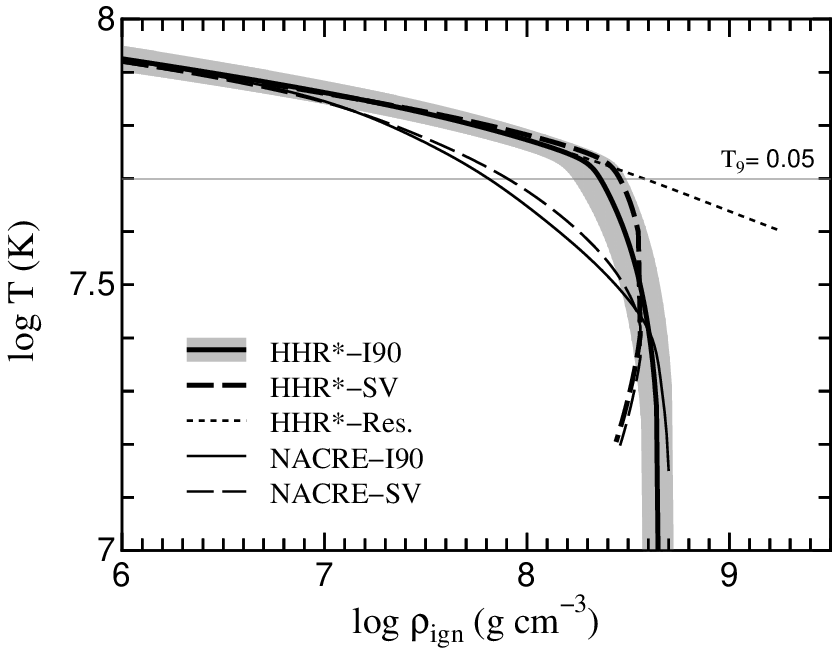}
  \end{center}
  \caption{\label{fig:ign}
    Ignition curves for helium burning in accreting white dwarfs.
    The bold solid and bold dashed curves are the results obtained from HHR$^\ast$ with I90 \cite{Ito90} and SV \cite{Sal69}.
    The uncertainties of $\tau_{\rm He}$ are displayed with the shade region, $\tau_{\rm He}=10^5$--10$^7$ yr \cite{Sar10}.
    The solid and dashed curves are calculated from NACRE.
    At $T_9=0.05$, helium burning is ignited in higher density than that expected with NACRE.
    The dotted curve is the result obtained only from the resonant rates.
  }
\end{figure}

The temperature dependence of $\rho_{\rm ign}$ is displayed in \fref{fig:ign}.
This figure illustrates an ignition point of accumulation starting from low densities and low temperatures, and it is used to see the impact of variation of the triple-$\alpha$ reaction rates in astrophysical applications.
The bold solid and bold dashed curves are the results obtained from the present recommended rates with I90 and SV.
The uncertainties in $\tau_{\rm He}$ are displayed with the gray shade, $\tau_{\rm He}=10^5$--10$^7$ yr \cite{Sar10}.
The solid and dashed curves are calculated from the NACRE rates.
The calculated density for HHR$^\ast$ is found to become insensitive to temperatures in $0.01 < T_9 < 0.05$, owing to the absence of the non-resonant sequential process between $\alpha+^8$Be.
The dotted curve is the result obtained only from the resonant term.
The non-resonant term, i.e. the direct process, is confirmed to play an important role in helium burning of accreting white dwarfs at $T_9 =0.01$, which seems to be consistent with \cite{Nom85,Lan86}.
In addition, I figure out that the difference between the ignition curves obtained from HHR$^\ast$ and NACRE is much smaller than that in the previous discussion \cite{Sar10}.
I therefore reckon that the astrophysical impact is not so large, although the present rates are reduced strongly around $T_9=0.05$.

The ignition curves, which are located at lower temperatures for a given ignition density, mean earlier ignition of helium burning.
Thus, NACRE might have facilitated earlier ignition.
Considering evolutionary tracks shown in figure~2 of \cite{Sar10}, I find that the present results affect the beginning of helium flash in two CO white dwarfs: initial mass $M_{\rm CO}=1.08$ M$_\odot$ with accretion rate $dM/dt=3\times10^{-9}$ M$_\odot$ yr$^{-1}$ and $M_{\rm CO}=1.28$ M$_\odot$ with $dM/dt=7\times10^{-10}$ M$_\odot$ yr$^{-1}$.
The intersection between the curves for NACRE and HHR$^\ast$ at log~$T \approx 7.42$ in the high density region corresponds to the unity at $T_9\approx0.026$ in \fref{fig:rates}(B).
Below this point, the understanding of white dwarfs with slower accretion is expected to remain intact because the difference from NACRE is within the uncertainties of ignition.

\section{Summary}\label{sec:6}

The present article has discussed the direct triple-$\alpha$ process and derived reaction rates by using the non-adiabatic Faddeev HHR$^\ast$ expansion method.
First, I have reviewed this theory and setup.
The main differences between HHR and HHR$^\ast$ seem to stem from the matching procedure for asymptotic wavefunctions (\sref{sec:2.4}) and Coulomb potentials in the asymptotic region (\sref{sec:2.2}), in addition to the adopted basis functions.
In \sref{sec:3.2}, I have introduced the three-body $S$-factors, and I have deduced the formula of reaction rates, attaching the resonant component.

In \sref{sec:4}, I have described the adopted potentials and the generated states of $^8$Be and $^{12}$C, and I have examined the available range $r_{3\alpha}$ of 3$\alpha$ potentials.
The CD potential, based on the DF model, reproduces the experimental phase shifts of $\alpha$+$\alpha$ elastic scattering and the $^8$Be(0$^+_1$) resonance.
In contrast, the nuclear potential of AB is shallower than that of CD, and it cannot reproduce the $l_x=4$ phase shifts.
The density distribution function of 0$^+_2$ in $^{12}$C is confirmed to have dual aspects: dilute $\alpha$+$^8$Be and spatially-shrunk structure.
The mixing ratio of this dilute clustering depends on $r_{3\alpha}$.
Whereas $r_{3\alpha}=3.46$ fm (CDa) and 4.0 fm (CDb) are recommended by CMF and ACF, $r_{3\alpha}=5.0$ fm (CDc) and 6.0 fm (CDd) seem to be preferable from \cite{Ots22,Toh01}.
In addition, I have confirmed that the 2$^+_1$ state has a spatially-shrunk configuration of an equilateral triangle with the almost same size as the ground state.
The resultant rms radii of 0$^+_2$ and 2$^+_1$ seem to be consistent with those from HHR, CMF and ACF.
The calculated charge radius of 0$^+_1$, $B(E2;0^+_2\rightarrow 2^+_1)$, $B(E2;2^+_1\rightarrow 0^+_1)$, $M(E0;0^+_2\rightarrow 0^+_1)$, and $\Gamma(0^+_2)$ are also concordant with the experimental values.

I have discussed the calculated reaction rates in sections~\ref{sec:5.1} and \ref{sec:5.2}, illustrating the calculated photo-disintegration cross sections of $^{12}$C(2$^+_1 \rightarrow 0^+$) and the corresponding $S$-factors.
At the same time, I have discussed the difference between adiabatic and non-adiabatic calculations.
The calculated photo-disintegration at off-resonant energies is estimated to be in 10$^{-15}$ -- 10$^{-3}$ pico-barn order of cross sections $\sigma_{\rm g}$ for $0.15 \le E \le 0.35$ MeV.
This is far below the values predicted by ACF and CMF.
Despite the large difference, the derived reaction rates are concordant with the NACRE rates for $0.08 \le T_9 \le 3$.
The difference in $\sigma_{\rm g}$ below $E = 0.20$ MeV can be seen in the rates for $T_9 \le 0.07$.
In comparison with the calculations, I have found that the triple-$\alpha$ reaction rates are reduced by a factor of 10$^{-4}$ at $T_9 = 0.05$, because of the accurate description for $^8$Be break-up.
The recommended rates are made from CDb.
The uncertainties of the rates have been estimated by examining the sensitivity to the 3$\alpha$ potentials.
The upper and lower limits are estimated from CDa and CDc within a factor of 2 for $T_9 \le 0.05$.
The rates obtained from AB are similar to those from CD.
Thus, I have confirmed that the reduction at $T_9=0.05$ does not stem from the adopted $\alpha$+$\alpha$ potentials.
To examine the long-range Coulomb couplings, I have also demonstrated the calculations without screening, and I have found that the Coulomb coupling effects are negligible for $\rho > 650$ fm at the energies corresponding to helium burning temperatures.

In \sref{sec:5.3}, I have converted the derived rates into REACLIB format, and I have shown that the present rates do not have the component of the non-resonant sequential process between $\alpha$+$^8$Be.
My calculated results therefore suggest that this component should be explicitly removed to update the rates in NACRE and REACLIB.
The current standard rates are based on a method with the particle approximation of $^8$Be, and they have the extra component due to the approximation.
In astrophysical applications with REACLIB, you can delete the records of the corresponding coefficients in \eref{eq:reaclib}, which are normally loaded from an external file.
The numerical reaction rates of the present study are available in the tabular form, in addition to the analytic form.

Finally, I have confirmed that the present rates do not lead to the drastic change in astrophysical applications.
In addition, I have confirmed that the direct process is important for helium burning in accreting white dwarfs.
Due to the reduction of rates at $T_9= 0.05$, the resulting ignition density has been found to become insensitive to temperatures in $0.01 < T_9 < 0.05$.
Further discussion about the astrophysical impacts will be made elsewhere.

\ack
I thank M.~Arnould, Y.~Sakuragi and Y.~Ohnita for drawing my attention to this subject and for their hospitality during my stay at Universit\'e Libre de Bruxelles and Osaka City University.
I am also grateful to Professor I.J. Thompson for his valuable comments.
This research did not receive any specific grant from funding agencies in the public, commercial, or not-for-profit sectors.

\appendix
\section{Convergence tests for basis parameters}\label{app:1}
The convergence property for basis parameters is complemented here.

\begin{figure}[t] 
  \begin{center}
    \includegraphics[width=0.6\linewidth]{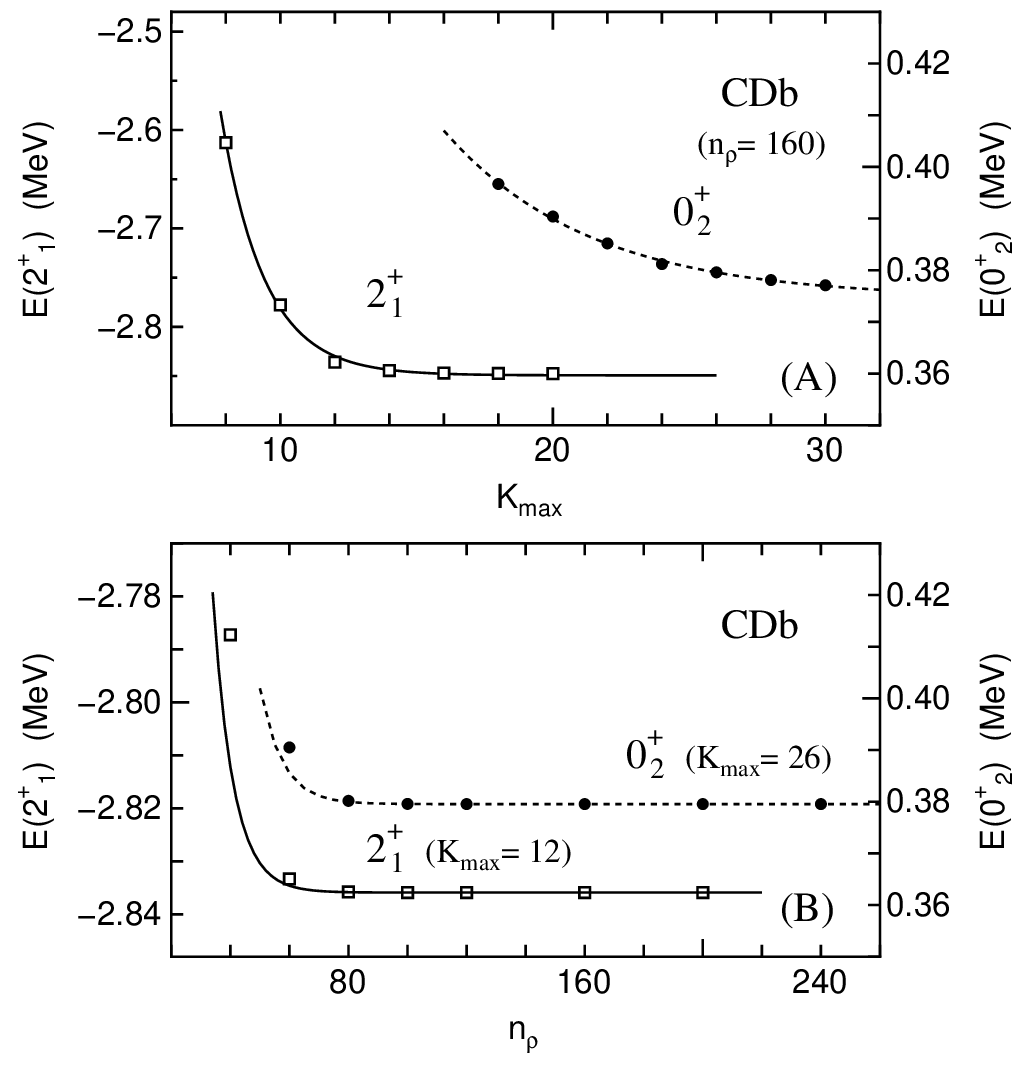}
  \end{center}
  \caption{\label{fig:cvg}
    (A) Dependence of $E(0^+_2)$ and $E(2^+_1)$ on $K_{\rm max}$.
    $K \le K_{\rm max}$ are included in \eref{eq:basis}.
    The solid circles and open squares are $E(0^+_2)$ and $E(2^+_1)$, respectively.
    The dotted and solid curves are the fits by $E(l^\pi_i) = e_0 + e_1\exp(-c_{\rm K}\,K_{\rm max})$.
    $e_0$, $e_1$, $c_{\rm K}$ are the coefficients of expansion.
    $n_\rho = 160$ and CDb are used.
    (B) Same but on $n_\rho$ of \eref{eq:varphi}.
    $K_{\rm max}=26$ ($K_{\rm max}=12$) is used for $0^+_2$ ($2^+_1$).
  }
\end{figure}

\Fref{fig:cvg}(A) shows the calculated $E(l^\pi_i)$ with the maximum hyperangular momentum $K_{\rm max}$.
$K \le K_{\rm max}$ are included in \eref{eq:basis}.
The solid circles and open squares represents $E(0^+_2)$ and $E(2^+_1)$, respectively.
From this figure, the convergence property of the present study is found to be similar to that of \cite{Ngu13}, i.e., $E(0^+_2)$ ($E(2^+_1)$) starts converging from $K_{\rm max} = 26$ ($K_{\rm max} = 12$).
As performed in \cite{Ngu13}, they are fitted by $E(l^\pi_i) = e_0 + e_1\exp(-c_{\rm K}\,K_{\rm max})$.
$e_0$, $e_1$, $c_{\rm K}$ are the coefficients of expansion, $c_{\rm K}>0$.
The resultant fits for 0$^+_2$ and 2$^+_1$ are displayed by the dotted and solid curves.
When $K_{\rm max}\rightarrow\infty$, this function is assumed to approach the converged value.
The differences from $E(l^\pi_i)$ extracted by CDb are therefore estimated to be 4.9 keV (1.3\%) at $K_{\rm max}=26$ for 0$^+_2$ and 13 keV (0.46\%) at $K_{\rm max}=12$ for 2$^+_1$.
\Fref{fig:cvg}(B) demonstrates a test for $n_\rho$ of \eref{eq:varphi}.
The dotted and solid curves are the fits in the similar expression to that in \fref{fig:cvg}(A).
The converged value is given by $n_\rho\rightarrow\infty$.
Accordingly, the differences are 10 meV for 0$^+_2$ and less than 1 meV for 2$^+_1$ at $n_\rho=160$.
$K_{\rm max}=26$ ($K_{\rm max}=12$) is used for $0^+_2$ ($2^+_1$).

$E(l^\pi_i)$ do not depend on the variation of $\nu$ in \eref{eq:gwf0} because $n_\rho$ is large.
However, $\nu$ is set in a certain range so as to make expansion efficiently because $\varphi^K_n(\rho)$ are dominated by a factor of $\exp[-(\nu\rho)^2/2]$ at the large $\nu\rho$, and plunge into zero.
This is for convenience of numerical processing, rather than $\nu$-dependence.
$\varphi^K_n(\rho)$ is a set of arbitrary orthonormal functions in theory.

\section{NACRE analytic expansion}\label{app:2}

The NACRE analytic expansion \cite{Nacre} is given, as follows:
\begin{eqnarray}
  \langle R_{3\alpha} \rangle_{\rm NACRE} &=&  R_{\alpha\alpha}\, R_{\alpha {\rm ^8Be}}\, F_T
  \nonumber\\
  &=& (R_{\alpha\alpha}^{\rm R}R_{\alpha {\rm ^8Be}}^{\rm R}
  +R_{\alpha\alpha}^{\rm R}R_{\alpha {\rm ^8Be}}^{\rm NR}
  +R_{\alpha\alpha}^{\rm NR}R_{\alpha {\rm ^8Be}}^{\rm R}
  +R_{\alpha\alpha}^{\rm NR}R_{\alpha {\rm ^8Be}}^{\rm NR}\,)\, F_T
  \nonumber\\
  &\equiv& R_{\rm DR}+ R_{\rm 8Be} + R_{\rm 12C} + R_{\rm NRs},
  \\
  &\approx& R_{\rm DR} +R_{\rm 8Be} +R_{\rm NRs},
  \label{eq:nacre-analytic}
\end{eqnarray}
where $R_{\alpha\alpha}=R_{\alpha\alpha}^{\rm R} +R_{\alpha\alpha}^{\rm NR}$ and $R_{\alpha {\rm ^8Be}}= R_{\alpha {\rm ^8Be}}^{\rm R}+R_{\alpha {\rm ^8Be}}^{\rm NR}$.
$R_{\alpha\alpha}$ and $R_{\alpha {\rm ^8Be}}$ are the reaction rates of $\alpha$+$\alpha$ and $\alpha$+$^8$Be, respectively.
The superscript R (NR) means the resonant (non-resonant) term.
$R_{\rm 12C}$ is negligible.
$F_T$ is given by
\begin{eqnarray}
  F_T &=&
  \left\{\begin{array}{ll}
    3.07\times10^{-16} \cdot (1-29.1 \,T_9+1308 \,T_9^2) & T_9\le0.03 \\
    3.44\times10^{-16} \cdot (1+0.0158 \,T_9^{-0.65})  & T_9 > 0.03 \\
  \end{array}\right.
\end{eqnarray}
$R_{\rm DR}$, $R_{\rm 8Be}$ and $R_{\rm NRs}$ are defined as
\begin{eqnarray}
  R_{\rm DR} &=& 7.96\times10^{7}\,T_9^{-3}\,F_T\,\exp(-4.392/T_9)
  \nonumber\\
           &+&1.53\times10^{10}\,T_9^{-3}\,F_T\,\exp(-21.36/T_9),
  \\
  R_{\rm 8Be} &=& 1.681\times10^{13} \,T_9^{-13/6} (1+5.47\,T_9+326\,T_9^2)
  \nonumber\\
  &\cdot&F_T \exp\Big[-1.054/T_9-23.57\,T_9^{-1/3}-(T_9/0.4)^2\Big],
  \\
  R_{\rm NRs} &=& 6.71\times10^{16}\,T_9^{-4/3}\,
  (1\!+\!79.97\,T_9\!+\!733.5\,T_9^2\!+\!2.429\times10^4\,T_9^3)
  \nonumber\\
  &\cdot&F_T \exp\Big[-37.06\,T_9^{-1/3} -50.69\,T_9^2\Big].
\end{eqnarray}
\Eref{eq:nacre-analytic} is displayed in \fref{fig:reaclib}(D).
In \fref{fig:rates} and \tref{tb:rates}, the present rates are compared with the values in the tabular form of NACRE.

\end{document}